\documentclass[%
 aip,
 amsmath,amssymb,
 reprint,%
]{Definitions/revtex4-1}

\usepackage{graphicx}
\usepackage{dcolumn}
\usepackage{bm}

\usepackage[utf8]{inputenc}
\usepackage[T1]{fontenc}
\usepackage{mathptmx}
\usepackage{etoolbox}
\usepackage{comment}
\usepackage{caption}
\DeclareMathOperator*{\Var}{Var}
\DeclareMathOperator*{\Cov}{Cov}
\DeclareCaptionOptionNoValue*{ljustified}{%
  \caption@setformat{plain}%
  \caption@setjustification{ljustified}%
}
\DeclareCaptionJustification{ljustified}{\ljustified}
\makeatletter
\newcommand\ljustified{%
  \let\\\@fillcr
  \leftskip\z@\@plus +0fil
  \rightskip\z@\@plus +0fil
  \parfillskip\z@\@plus 1fil\relax
}
\makeatother
\usepackage{enumerate}

\makeatletter
\def\@email#1#2{%
 \endgroup
 \patchcmd{\titleblock@produce}
  {\frontmatter@RRAPformat}
  {\frontmatter@RRAPformat{\produce@RRAP{*#1\href{mailto:#2}{#2}}}\frontmatter@RRAPformat}
  {}{}
}%
\makeatother
\begin{document}

\preprint{AIP/123-QED}

\title[Information-theoretical analysis of statistical measures for multiscale dynamics]{Information-theoretical analysis of statistical measures for multiscale dynamics}
\author{Naoki Asuke}
\email{jintengzhizi196@gmail.com}
\affiliation{ 
Department of Information Physics and Computing, Graduate School of Information Science and Technology, The University of Tokyo, 7-3-1 Hongo, Bunkyo-ku, Tokyo 113-8656, Japan
}
\author{Tomoki Yamagami}%
\affiliation{ 
Department of Information Physics and Computing, Graduate School of Information Science and Technology, The University of Tokyo, 7-3-1 Hongo, Bunkyo-ku, Tokyo 113-8656, Japan
}%

\author{Takatomo Mihana}%
\affiliation{ 
Department of Information Physics and Computing, Graduate School of Information Science and Technology, The University of Tokyo, 7-3-1 Hongo, Bunkyo-ku, Tokyo 113-8656, Japan
}%

\author{Andr\'{e} R\"{o}hm}%
\affiliation{ 
Department of Information Physics and Computing, Graduate School of Information Science and Technology, The University of Tokyo, 7-3-1 Hongo, Bunkyo-ku, Tokyo 113-8656, Japan
}%

\author{Ryoichi Horisaki}%
\affiliation{ 
Department of Information Physics and Computing, Graduate School of Information Science and Technology, The University of Tokyo, 7-3-1 Hongo, Bunkyo-ku, Tokyo 113-8656, Japan
}%

\author{Makoto Naruse}%
\affiliation{ 
Department of Information Physics and Computing, Graduate School of Information Science and Technology, The University of Tokyo, 7-3-1 Hongo, Bunkyo-ku, Tokyo 113-8656, Japan
}%


\date{\today}

\begin{abstract}
Multiscale entropy (MSE) has been widely used to examine nonlinear systems involving multiple time scales, such as biological and economic systems. 
Conversely, Allan variance has been used to evaluate the stability of oscillators, such as clocks and lasers, ranging from short to long time scales. 
Although these two statistical measures were developed independently for different purposes in different fields in the literature, their interest is to examine multiscale temporal structures of physical phenomena under study. 
We show that, from an information-theoretical perspective, they share some foundations and exhibit similar tendencies. 
We experimentally confirmed that similar properties of the MSE and Allan variance can be observed in low-frequency fluctuations (LFF) in chaotic lasers and physiological heartbeat data. 
Furthermore, we calculated the condition under which this consistency between the MSE and Allan variance exists, which is related to certain conditional probabilities. 
Heuristically, physical systems in nature including the aforementioned LFF and heartbeat data mostly satisfy this condition, and hence the MSE and Allan variance demonstrate similar properties. 
As a counterexample, an artificially constructed random sequence is demonstrated, for which the MSE and Allan variance exhibit different trends. 
\end{abstract}

\maketitle

\section{\label{sec:intro}Introduction}
Multiscale entropy (MSE) \cite{PhysRevLett.89.068102} has been used widely to evaluate nonlinear systems that involve multiple time scales in biology,\cite{CATARINO20112375,PhysRevE.71.021906}  economics,\cite{MARTINA2011936} transportation,\cite{doi:10.1142/S012918311350006X} and other fields.
Meanwhile, Allan variance \cite{AV} has been used to evaluate the stability of oscillators,\cite{AV,crystal,laserAV} such as atomic clocks or lasers, over many time scales.
Although these two statistical measures were developed independently in different domains, the MSE and its variants\cite{e17053110} and Allan variance have a similar objective: to characterize dynamical systems containing multiple time scales. 

However, the relevance and differences between the MSE and Allan variance have yet to be examined in the literature. 
In this study, we discuss the similarities and differences between the MSE and Allan variance from an information-theoretical perspective and from the viewpoint of computational cost, and reveal the mechanisms behind them. 
We experimentally validated the similar tendencies of the MSE and Allan variance observed in low-frequency fluctuations (LFF) in chaotic lasers that are numerically simulated by the Lang--Kobayashi equations,\cite{LK1980} and physiological heartbeat data available in the public domain.\cite{PhysioNet,healthy,BAIM1986661,10.1093/europace/eum096} 
Furthermore, we present the underlying mechanism behind the consistency between the MSE and Allan variance based on conditional probabilities. 
We artificially constructed a random sequence that violated this condition, leading to a case exhibiting inconsistent results between the MSE and Allan variance.

As discussed in detail below, the LFF and heartbeat contain both slow and fast dynamics; they represent typical physical phenomena with multiple time scales. 
Furthermore, recent studies on physics-based computing and communications, such as reservoir computing,\cite{Larger:12} laser-chaos-based secure communication,\cite{Rogister:01,ChaosComm2} and laser networks for solving reinforcement learning problems,\cite{Mihana:20} work across multiple time scales; 
hence, understanding fundamental attributes in multiple time scales through statistical measures, such as the MSE and Allan variance, is essential for furthering our comprehension.

The remainder of this paper is organized as follows. 
We review the Allan variance and MSE in Sections~\ref{sec:AV} and~\ref{sec:MSE}, respectively. 
Section~\ref{sec:simulation} examines the MSE and Allan variance for several time series, namely noise, RR interval,\cite{BA85044686} and laser chaos. 
A similar tendency of the MSE and Allan variance is shown for each time series. Section~\ref{sec:discussion} discusses the mechanism behind the similarity between the two measures. 
Additionally, the cause of the differences in behavior is discussed. Section~\ref{sec:conclusion} concludes the paper.

\section{Theory}
\subsection{\label{sec:AV} Allan variance}
Coarse-graining of a time series $s(t)~(t\in \{1,2,\cdots,N\})$ with a scale factor $\tau_A$ refers to the operation that obtains another time series $\bar{s}^{(\tau_A)}(l)~(l\in \{1,2,\cdots,\lfloor N/\tau_A\rfloor\})$ as follows:
\begin{align}
    \bar{s}^{(\tau_A)}(l) &= \dfrac{1}{\tau_A} \sum_{t=(l-1)\tau_A+1}^{l\tau_A} s(t). \label{eq:CG1}
\end{align}
Here, $\lfloor\,\cdot\,\rfloor$ denotes floor function.
The time series $\bar{s}^{(\tau_A)}(l)$ is called a coarse-grained time series.
This operation obtains a new point by taking the average of every $\tau_A$ point.
Using coarse-graining, the Allan variance\cite{AV} of a time series $s(t)$ with scale factor $\tau_A$ is defined as
\begin{align}
    \sigma_s^2(\tau_A) &= \dfrac{1}{2\lfloor N/\tau_A \rfloor}\sum_{l=2}^{\lfloor N/\tau_A \rfloor}\left(\Delta\bar{s}^{(\tau_A)}(l)\right)^2, \label{eq:AV_1} 
\end{align}
where 
\begin{align}
\Delta\bar{s}^{(\tau_A)}(l) &= \bar{s}^{(\tau_A)}(l) - \bar{s}^{(\tau_A)}(l-1). \label{eq:AV_2}
\end{align}
The Allan variance considers the variance of the difference between successive points in the coarse-grained time series under study, with respect to the time scale given by the coarse-graining time $\tau_A$.

Meanwhile, it is known that the Allan variance can be expressed using the power spectral density $S(f)$ of the signal $s(t)$ as follows, under the assumption that $s(t)$ is stationary and ergodic\cite{5570702}:
\begin{align}
    \lim_{N\to\infty}\sigma^2_s(\tau_A) &= 2\int_0^{\infty} S(f)\dfrac{\sin^4(\pi f \tau_A)}{(\pi f \tau_A)^2} df. \label{eq:AV_PSD}
\end{align}
Using Eq.~(\ref{eq:AV_PSD}), 
for example, 
the Allan variance of white noise is as follows:
\begin{align}
    2\int_0^\infty h_0\dfrac{\sin^4(\pi f \tau_A)}{(\pi f \tau_A)^2} df = \dfrac{h_0}{2\tau_A}. \label{eq:PhaseStabilityExample}
\end{align}
Similarly, for 
$1/f$ noise, the Allan variance is
\begin{align}
    2\int_0^\infty \dfrac{h_{-1}}{f}\dfrac{\sin^4(\pi f \tau_A)}{(\pi f \tau_A)^2} df = 2h_{-1}\ln 2. \label{eq:PhaseStabilityExample2}
\end{align}
Here $h_0$ and $h_{-1}$ represent the intensities of the white and $1/f$ components, respectively.
We can then estimate $h_0$ and $h_{-1}$ from the Allan variance with various $\tau_A$.
In this way, the Allan variance has been used to evaluate the variability characteristics of time-series data.

\subsection{\label{sec:MSE} Multiscale entropy (MSE)}
Before discussing multiscale entropy,\cite{PhysRevLett.89.068102} it is necessary to introduce original sample entropy (SaEn). \cite{doi:10.1152/ajpheart.2000.278.6.H2039} 
The theoretical background and examples of SaEn were described in detail by Richman and Moorman.\cite{e21060541} We will use their definition of SaEn.

In short, SaEn quantifies the regularity of a time series. 
SaEn has several parameters: the embedding dimension $m$, tolerance $r$ and length of the time series $N$.
The SaEn of a time series $s(t)$ is defined with the correlation integral $C^m(r)$ as follows:
\begin{align}
    SaEn(m,r,N) = -\log \dfrac{C^{m+1}(r)}{C^m(r)}. \label{eq:SaEn}
\end{align}
Several steps must be taken to define and compute $C^m(r)$.
First, the embedded vector series $\bm{v}^m(t)$ is constructed as follows:
\begin{align}
    \bm{v}^m(t) &= [s(t),s(t+1),\cdots,s(t+m-1)]^T. \label{eq:embed}
\end{align}
Please note that the length of the vector series defined as Eq.~(\ref{eq:embed}) is $N-m+1$ because the $m$-th component of $\bm{v}^m(N-m+1)$ is $s(N)$.
Second, for each embedded vector, $C_i^m(r)$ and $C_i^{m+1}(r)$ are defined as follows:
\begin{align}
        C_i^m(r) &= \dfrac{1}{N-m-1}\sum_{\substack{j=1\\j\neq i}}^{N-m} u(r-d_{\rm cheb}(\bm{v}^m(i),\bm{v}^m(j))), \label{eq:Cmi} \\
        C_i^{m+1}(r) &= \dfrac{1}{N-m-1}\sum_{\substack{j=1\\j\neq i}}^{N-m} u(r-d_{\rm cheb}(\bm{v}^{m+1}(i),\bm{v}^{m+1}(j))).
\end{align}
Here, $u$ denotes the unit step function, defined as
\begin{align}
    u(x) = 
    \begin{cases}
        0 & (x<0)\\
        1 & (x \geq 0)
    \end{cases}.
\end{align}
$d_{\rm cheb}$ is the Chebyshev distance and $d_{\rm cheb}(\bm{v}^m(i),\bm{v}^m(j))$ is defined as follows:
\begin{align}
    d_{\rm cheb}(\bm{v}^m(i),\bm{v}^m(j)) &= \max_{k\in \{0,1,\cdots,m-1\}}|s(i+k)-s(j+k)|.
\end{align}
Please note that the maximum $j$ is not $N-m+1$ but $N-m$ in Eq.~(\ref{eq:Cmi}) because we also calculate $C_i^{m+1}(r)$ to compute SaEn and the total number of embedded vectors in the $m+1$ dimension is $N-m$.
Intuitively, the definition of $C^m_i(r)$ is the probability of a randomly chosen $\bm{v}^m(j)~(j\neq i)$ satisfying
\begin{align}
    d_{{\rm cheb}}(\bm{v}^m(i),\bm{v}^m(j)) &< r.
\end{align}
It is noteworthy that, by the definition of the Chebyshev distance, the following proposition holds:
\begin{align}\begin{split}\label{iff:cheb}
    &d_{{\rm cheb}}(\bm{v}^m(i),\bm{v}^m(j)) < r \\
    \mbox{iff}\ \, &\forall k\in \{0,1,\cdots,m-1\},\ |s(i+k)-s(j+k)|<r.
\end{split}\end{align}
This proposition plays an important role in Sec.~\ref{sec:CIandAV}.

From the definition, $C^m_i(r)$ can be regarded as the probability that a randomly chosen $\bm{v}^m(j)$ is a neighbor of the embedded vector $\bm{v}^m(i)$. 
In other words, it is the likelihood of patterns $\left[s(j),s(j+1),\cdots s(j+m-1) \right]^T$ which are considered repetitions of the pattern $\left[s(i),s(i+1),\cdots, s(i+m-1) \right]^T$ under the tolerance $r$. 
With $C_i^m(r)$, the correlation integral $C^m(r)$ is defined as 
\begin{align}
    C^m(r) &= \dfrac{1}{N-m}\sum_{i=1}^{N-m}C_i^m(r). \label{eq:Cm}
\end{align}
Intuitively, $C^m(r)$ is the average of $C_i^m(r)$; therefore, we can regard $C^m(r)$ as the probability that randomly chosen vectors $\bm{v}^m(i)$ and $\bm{v}^m(j)~(i\neq j)$ are close.
A large $C^m(r)$ indicates that time series $s(t)$ contains many repeating structures of length $m$. 
SaEn then quantifies whether, when $m$ consecutive points in the time series are considered repeated, the $(m+1)$-th point is also considered repeated. 

In other words, it quantifies whether the $(m+1)$-th point is predictable by looking at the $m$ preceding points.
SaEn can be applied to real-world data without assuming any model governing the time series. 

The MSE is defined as the SaEn of a coarse-grained time series and
can be plotted as a function of $\tau_A$, where the same $r$ is used for every $\tau_A$. 
The MSE was proposed to quantify complexity at various time scales. $1/f$ noise has a long-time correlation; therefore, it is considered more complex than white noise. 
However, SaEn assigns the maximum value to white noise. 
In contrast, the MSE of white noise monotonically decreases with increasing $\tau_A$, whereas the MSE of $1/f$ noise is constant,\cite{PhysRevLett.89.068102} i.e., $1/f$ noise has more complexity than white noise over a longer time scale.

As introduced in Secs.~\ref{sec:AV} and \ref{sec:MSE}, the definitions of the MSE and Allan variance are apparently completely different. 
In this study, we demonstrate that the MSE and Allan variance exhibit similar $\tau_A$ dependency, and reveal the underlying mechanism from an information-theoretical perspective.
Note that for $1/f$ noise, it has already been shown theoretically that both the Allan variance and MSE are constants, independent of $\tau_A$.

\section{\label{sec:simulation} Comparison of the MSE and Allan variance}
To compare the behavior of the MSE and the correlation integral $C^m(r)$ to that of the Allan variance, they were calculated for the three types of signals described later. 
$m$ was fixed to $2$, and $r$ was fixed at $0.15 \times SD(s)$,\cite{PhysRevLett.89.068102} where $SD(s)$ denotes the standard deviation of the signal $s(t)$.

\subsection{\label{sec:noise} Noise}
\begin{figure}[htbp] 
\centering
\includegraphics[width=\columnwidth]{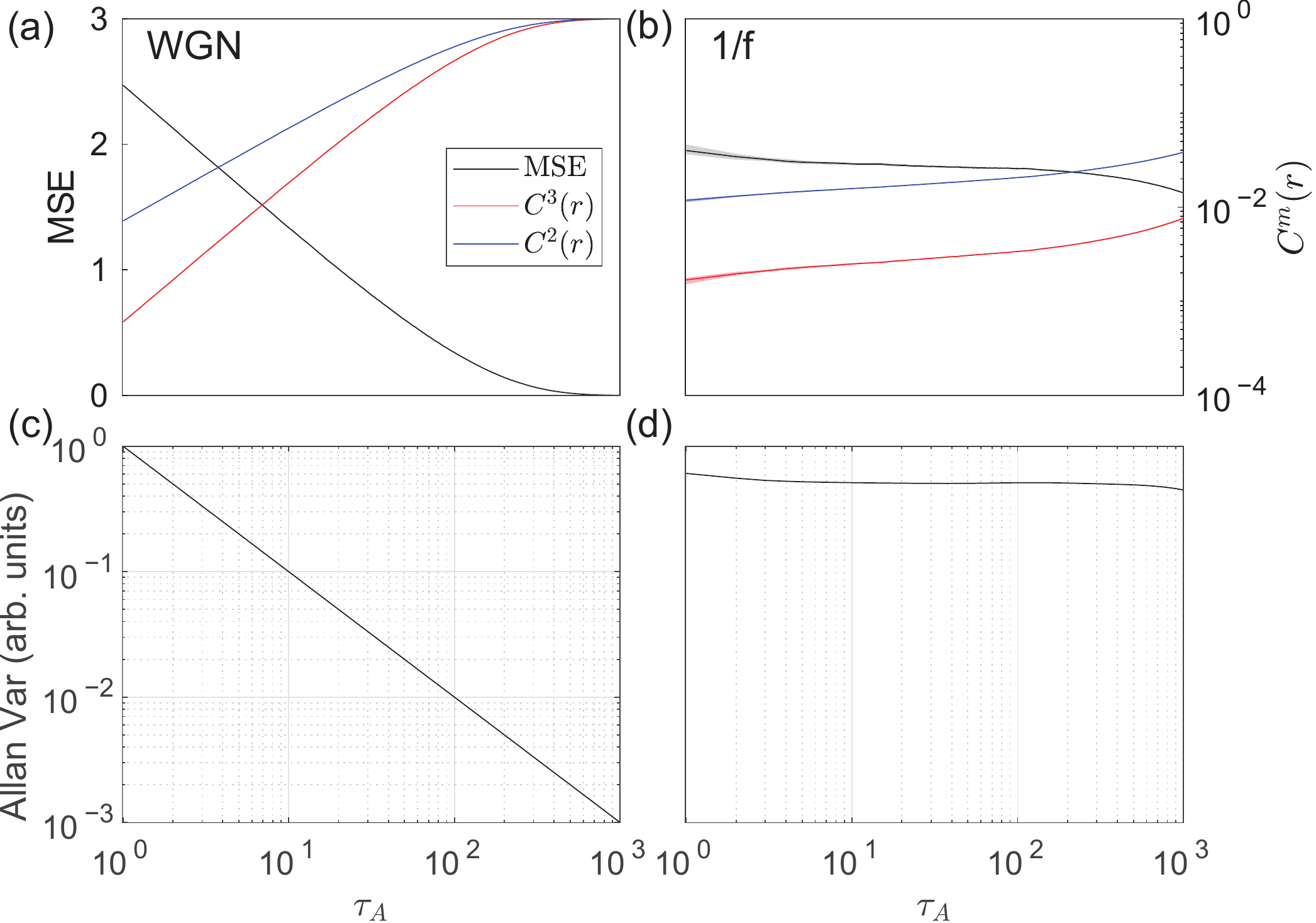}
\captionsetup{singlelinecheck=off,justification=ljustified}
\caption{
Comparison of the MSE and Allan variance for WGN and $1/f$ noise. 
(a,b) The MSE and $C^m(r)$ of WGN and $1/f$ noise, respectively. 
The black, red, and blue curves represent the MSE, $C^3(r)$ and $C^2(r)$, respectively. 
Each curve is the average of independent results from 30 trials, and the filled area around the curve represents the first and third quartile. 
The variance of the computed results is so slight that it is difficult to observe the painted areas.
(c,d) The Allan variance of WGN and $1/f$ noise, respectively. 
The Allan variance plots and $C^m(r)$ plots are logarithmic on both axes to observe the $1/\tau_A$ dependency of WGN clearly, while the y-axis is linear for the MSE plots, because the MSE is defined with the logarithm of $C^m(r)$. }
\label{fig:noise}
\end{figure}

Thirty temporal waveforms of white Gaussian noise (WGN) and $1/f$ noise containing $10^7$ points each were generated numerically, and the MSE and Allan variance were calculated up to $\tau_A = 10^3$ points. 
We fixed the length of the coarse-grained time series to $10^4$ points for the MSE calculation owing to limitations of our computing environment, whereas the entire time series was used for the Allan variance calculation. 
The results are presented in Fig.~\ref{fig:noise}. 

The Allan variance and MSE or $-C^m(r)$ show similar trends with regard to $\tau_A$, with appropriate scaling.
Please note that the scales of the y-axes in $C^m(r)$ plots and the Allan variance plots are identical (logarithmic), while the y-axes in the MSE plots are linear. Because the MSE is defined by the logarithm of $C^m(r)$, it is natural to plot it linearly when $C^m(r)$ is plotted logarithmically.

Figures~\ref{fig:noise}(a) and (c) show the MSE and Allan variance of WGN, respectively, where both decrease monotonically as a function of $\tau_A$. This trend agrees well up to $\tau_A = 10^2$ points. 
For $\tau_A > 10^2$ points, the Allan variance still shows $1/\tau_A$ dependency, as indicated by Eq.~(\ref{eq:PhaseStabilityExample}), whereas the MSE saturates at 0. 
In this region, $C^m(r)$ is almost 1, i.e., almost all of the embedded vectors are neighbors of each other. This saturation effect is one of the differences between the MSE and Allan variance. 

Figures~\ref{fig:noise}(b) and (d) show the MSE and Allan variance of $1/f$ noise, respectively. 
Theoretically, both are predicted to be independent of $\tau_A$, whereas the simulation results show some $\tau_A$ dependency. 
This may be a result of the difficulty in properly generating $1/f$ noise on scales from $\tau_A$ = $1$ to $10^3$, however it is noteworthy that the MSE, $-C^m(r)$ and Allan variance exhibit similar curves as a function of $\tau_A$.

\subsection{\label{sec:bio} Physiological signal}
Costa \textit{et~al}.\cite{PhysRevLett.89.068102} introduced the MSE to demonstrate that physiological signals observed in young, healthy individuals exhibit versatile signal levels across a variety of time scales, whereas unhealthy subjects do not. 
In this section, we compare the MSE and Allan variance of the RR interval\cite{BA85044686} series, which are sequences of intervals between adjacent R waves in electro-cardiograms, for healthy subjects and those with congestive heart failure (CHF) or atrial fibrillation (AF), following Costa \textit{et~al}.
Each dataset was obtained from PhysioNet.\cite{PhysioNet,healthy,BAIM1986661,10.1093/europace/eum096} 

\begin{figure*}[htbp] 
\centering
\includegraphics[width=1.95\columnwidth]{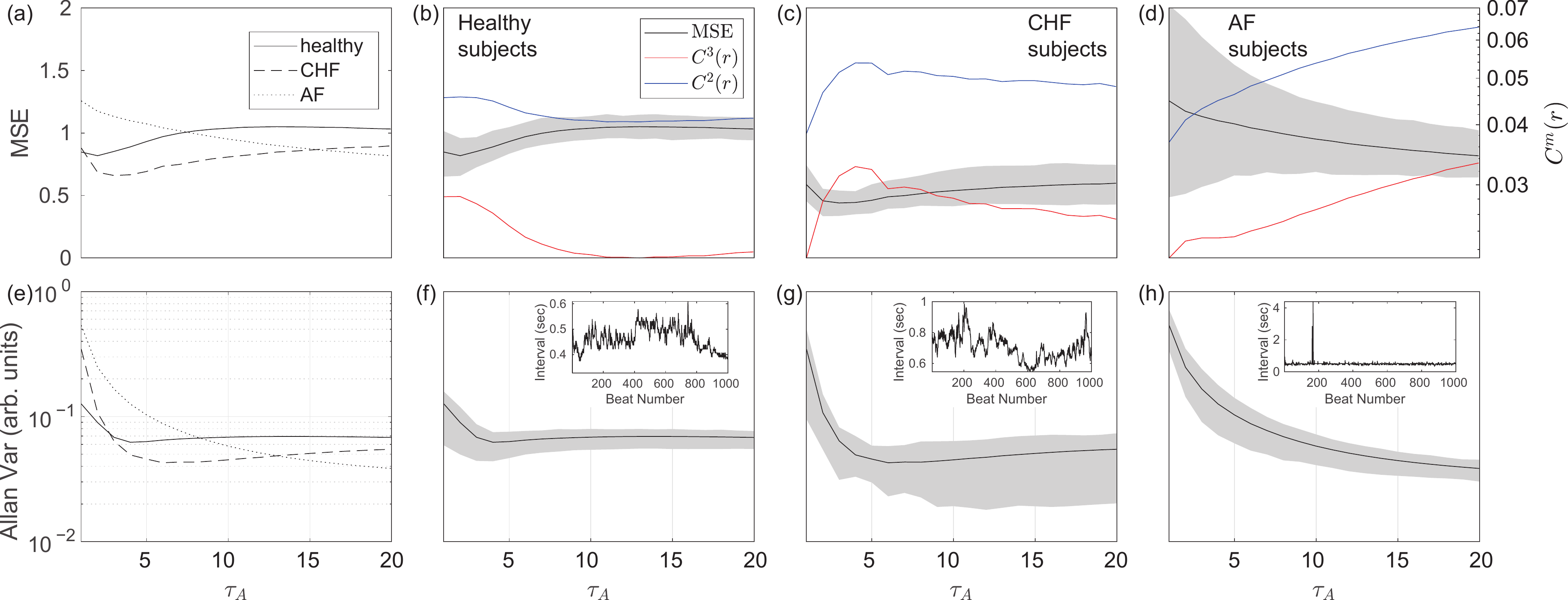}
\captionsetup{singlelinecheck=off,justification=ljustified}
\caption{
Comparison of the MSE and Allan variance for the RR interval time series. 
(a, e) Overlaid figure of the MSE and Allan variance of the RR interval time series from healthy subjects and subjects with CHF or AF, respectively.
(b,c,d) The MSE and $C^m(r)$ of the RR interval time series for healthy subject, CHF, and AF, respectively. 
The black, red, and blue curves represent the MSE, $C^3(r)$ and $C^2(r)$, respectively. 
Each curve is the average of independent results from 147 subjects for healthy, 29 for CHF, and 84 for AF, and the filled area around the MSE curve represents the first and third quartile. The first and third quartiles of $C^m(r)$ are omitted for legibility.
(f,g,h) Allan variance of the corresponding time series. The filled area around the curve represents the first and third quartile.
Each dataset was obtained from PhysioNet.\cite{PhysioNet,healthy,BAIM1986661,10.1093/europace/eum096} 
}
\label{fig:RR}
\end{figure*}

The lengths of the time series varied, but the entirety of each time series was used in the calculations, because adjusting to the shorter ones would reduce the accuracy of the estimation.
To compare the Allan variance, the standard deviation of each time series was normalized to 1. 
This normalization is consistent with the MSE calculation, because $r$ was fixed at $0.15\times SD(s)$.
The MSE and Allan variance for each dataset is shown in Fig.~\ref{fig:RR}.
The Allan variance, MSE and $-C^m(r)$ show trends that are similar except for the location of the peaks.

Figures~\ref{fig:RR}(a) and (e) respectively show the MSE and Allan variance overlaid for all three cases.
Figure~\ref{fig:RR}(a) shows that the RR interval for healthy subjects is the most complex on long time scales, which is consistent with the results of the previous study.\cite{PhysRevLett.89.068102} 
As shown in Fig.~\ref{fig:RR}(e), the ordering of the different classes of subjects for the Allan variance and MSE show a similar trend. 
However, the exact scale factor $\tau_A$ at which each curve intersects differs from that of the MSE. 
It seems plausible that the analysis of RR interval time series in general could be performed using Allan variance instead of the MSE.

\subsection{\label{sec:chaos} Laser chaos}
The MSE and Allan variance of the noise time series exhibit monotonic properties. 
However, the MSE and Allan variance of physiological signals show more complex properties, but the underlying dynamics are not completely known, which makes further analysis difficult. 

\begin{figure}[htbp]   
\centering
\includegraphics[width=\columnwidth]{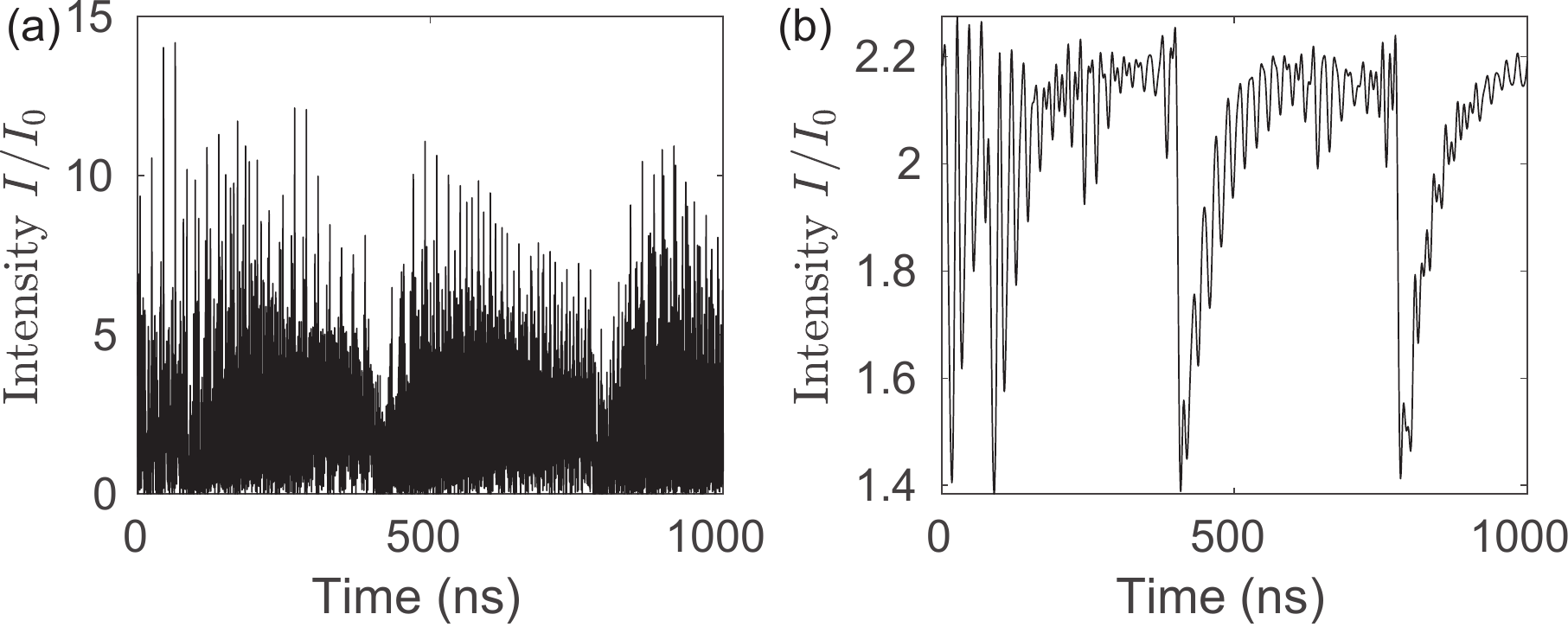}
\captionsetup{singlelinecheck=off,justification=ljustified}
\caption{
Example of an LFF time series. 
The intensity is normalized to that without optical feedback. 
(a) Original time series. 
(b) Time series obtained by applying an ideal low-pass filter with a cutoff frequency of 100 MHz to the time series in (a). 
Sudden dropouts with gradual recovery are observed.
}
\label{fig:LFF_example}
\end{figure}

\begin{figure}[htbp]  
\centering
\includegraphics[width=\columnwidth]{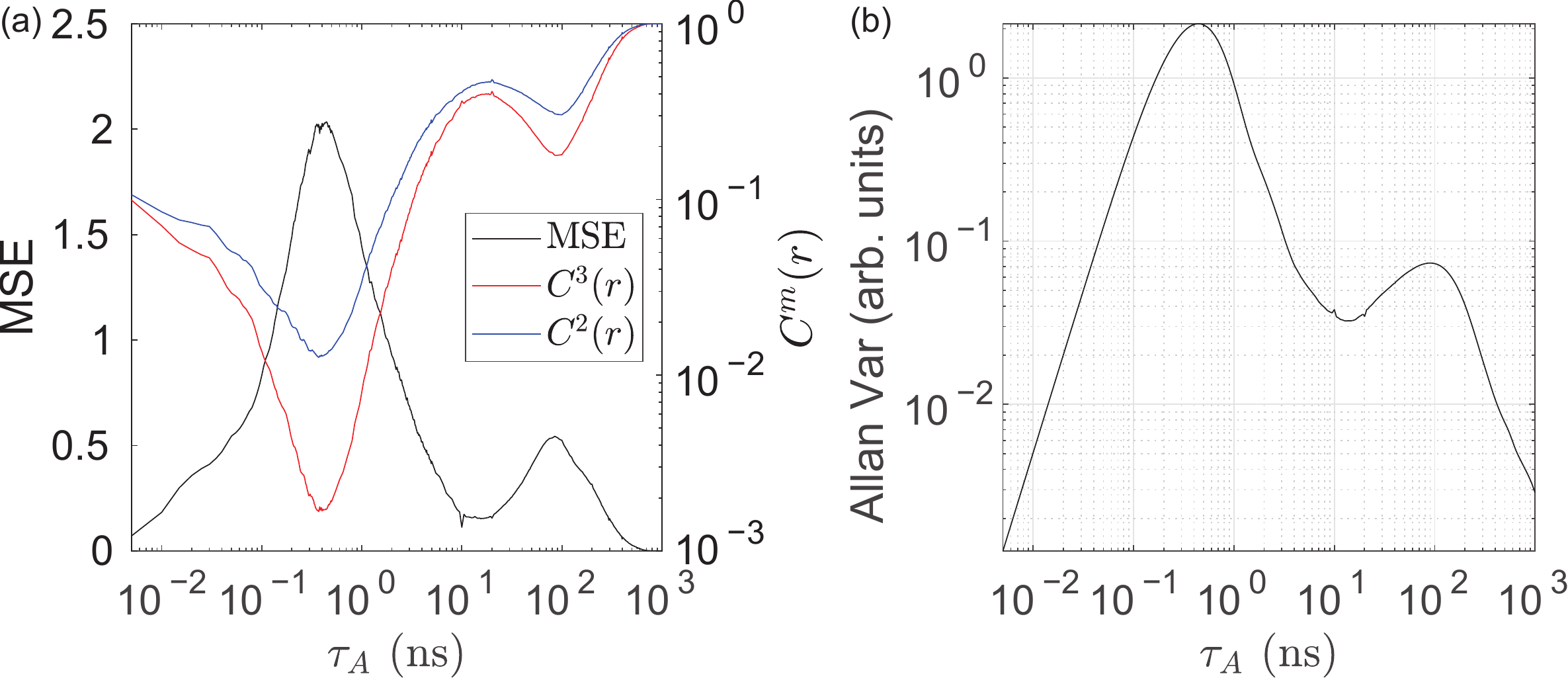}
\captionsetup{singlelinecheck=off,justification=ljustified}
\caption{
Comparison of the MSE and Allan variance for an LFF time series. 
(a) The MSE and $C^m(r)$. 
The black, red, and blue curves represent the MSE, $C^3(r)$ and $C^2(r)$, respectively. 
(b) Allan variance.
Note that the MSE, $-C^m(r)$, and the Allan variance exhibits similar $\tau_A$ dependencies.
}
\label{fig:LFF}
\end{figure}

As an example of a time series that includes multiple time scales and for which the underlying dynamics are known, we discuss a phenomenon called LFF\cite{Uchida12} exhibited by a semiconductor laser with optical feedback. 
The Lang--Kobayashi equations,\cite{LK1980} a set of model equations, was used to generate the time series. 
An example of a time series is shown in Fig.~\ref{fig:LFF_example}, where fast chaotic oscillations on the order of GHz coexist with irregular intensity dropouts and recovery on the order of MHz. 
The results are presented in Fig.~\ref{fig:LFF}. 
As the simulation step and time scale of the dynamics are important, $\tau_A$ was converted to time.

Figure~\ref{fig:LFF} shows that the MSE, $C^m(r)$ and Allan variance all capture the dynamics of fast oscillations on the order of GHz, corresponding to the fast peak in the MSE and Allan variance, and dropouts on the order of MHz, corresponding to the slow peak.


\section{\label{sec:discussion} Information theoretical connections between the MSE and Allan variance}

\subsection{\label{sec:CIandAV}$C^m(r)$ and Allan variance}
In Sec.~\ref{sec:simulation}, we observe that $-C^m(r)$ exhibits behavior similar to that of the Allan variance. 
In this section, we discuss the underlying mechanisms. 

\subsubsection{$C^m(r)$ decomposition \label{sec:CIdec}}
First, we consider decomposing $C^m(r)$ by the difference in the indices of the embedding vectors. 
Let $M = \lfloor N/\tau_A\rfloor$ be the length of the coarse-grained time series, and
$D^m(i,j)$ be $d_{{\rm cheb}}(\bm{v}^m(i),\bm{v}^m(j))$. Considering the expectation and symmetry of $i$ and $j$, and assuming $s(t)$ or its first-order difference $s(t)-s(t-1)$ is strongly stationary,
\begin{align}
    \langle C^m(r)\rangle &= \dfrac{\sum_{i=1}^{M-m}\sum_{\substack{j=1\\j\neq i}}^{M-m}\langle u \left(r - D^m(i,j)\right)\rangle}{(M-m-1)(M-m)} \label{eq:expect}\\
    &= \dfrac{\sum_{i=1}^{M-m}\sum_{\substack{j=1\\j\neq i}}^{M-m}\Pr\left(D^m(i,j)<r\right)}{(M-m-1)(M-m)} \label{eq:ij}\\
    &= \dfrac{\sum_{i=1}^{M-m}\sum_{j=i+1}^{M-m}2\Pr\left(D^m(i,j)<r\right)}{(M-m-1)(M-m)} \label{eq:ij_upper}\\
    &= \dfrac{\sum_{k=1}^{M-m-1}(M-m-k)\Pr(D^m(1,1+k)<r)}{\binom{M-m}{2}}. \label{eq:i-j_decomposition}
\end{align}
First we clarify the meaning of expectation. In the following, we regard the time series $s(t),~t\in\{1,2,\cdots,N\}$ under study as a sequence of random variables following a specific probability distribution function (PDF). In this sense, $C^m(r)$ and $u(r-D^m(i,j))$ are also random variables, and thus each of the possible values of $C^m(r)$ and $u(r-D^m(i,j))$ has some probability of occurrence. The expectation is the weighted average of the likelihood of every potential value.

Here we explain the transformation of the equations from (\ref{eq:expect}) to (\ref{eq:i-j_decomposition}) line by line.
The only change made from Eq.~(\ref{eq:expect}) to (\ref{eq:ij}) is that the expectation term $\langle u \left(r - D^m(i,j)\right)\rangle$ is changed to the probability term $\Pr\left(D^m(i,j)<r\right)$.
We assume that a random variable sequence $s(t),~t\in\{1,2,\cdots,N\}$ follows a PDF, and calculate $u \left(r - D^m(i,j)\right)$. As previously mentioned, $u \left(r - D^m(i,j)\right)$ is also a random variable that is either 0 or 1 with the following probability:
\begin{align}
    u \left(r - D^m(i,j)\right) = 
    \begin{cases}
        0 & (1-\Pr\left(D^m(i,j)<r\right)) \\
        1 & (\Pr\left(D^m(i,j)<r\right))
    \end{cases}.
\end{align}
Consequently,
\begin{align}
    \langle u \left(r - D^m(i,j)\right)\rangle = \Pr\left(D^m(i,j)<r\right)
\end{align}
holds.

From Eq.~(\ref{eq:ij}) to (\ref{eq:ij_upper}), the factor $2$ is added to the numerator, and the range of  summation over $j$ is restricted to $j>i$. Here, we use the fact that $D^m(i,j) = D^m(j,i)$.
From Eq.~(\ref{eq:ij_upper}) to (\ref{eq:i-j_decomposition}), we must assume that
 $\Pr\left(D^m(i,j)<r\right)$ depends only on the difference in the indices $k=j-i$. 
 For example, $\Pr\left(D^m(i+1,j+1)<r\right)$ is the same as $\Pr\left(D^m(i,j)<r\right)$.
This assumption is satisfied when the original signal $s(t)$ or its first-order difference $s(t)-s(t-1)$ is strongly stationary. This is a reasonable assumption when studying a time series from a dynamical system.
The factor $(M-m-k)$ in Eq.~(\ref{eq:i-j_decomposition}) comes from the fact that, for each $k$, the number of pairs $(i,j)$ satisfying $k=j-i$ is $(M-m-k)$. For example, when $k=1$, the pairs are $(1,2),(2,3),\cdots,(M-m-1,M-m)$, as the total number of embedded vectors is $M-m$.

The probability in Eq.~(\ref{eq:i-j_decomposition}) can be decomposed further by considering conditional probabilities.
Using the proposition~(\ref{iff:cheb}),
\begin{align}
    &\Pr(D^m(1,1+k)<r) \notag \\
    =\,&\Pr(d_{\rm cheb}(\bm{v}(1),\bm{v}(1+k))<r) \notag \\
    =\,&\Pr(|s(1)-s(1+k)|<r\land|s(2)-s(2+k)|<r\notag\\
    &\land\cdots\land |s(m)-s(m+k)|<r) \notag \\
    =\,&\Pr(D^1(1,1+k)<r\land D^1(2,2+k)<r\notag\\
    &\land\cdots\land D^1(m,m+k)<r) \label{eq:P_decompose}
\end{align}
holds. The probability of a product event, such as Eq.~(\ref{eq:P_decompose}), can be represented as a product of conditional probabilities. For example, the probability that propositions $A$, $B$ and $C$ hold simultaneously is
\begin{align}
    \Pr(A\land B\land C) &= \Pr(B\land C\mid A)\Pr(A) \notag\\
    &= \Pr(C\mid A\land B)\Pr(B\mid A)\Pr(A).
\end{align}
In the same way,
\begin{align}
    &\Pr(D^m(1,1+k)<r) \notag \\
    = &\Pr(D^1(1,1+k)<r)\notag\\&\prod_{i=2}^{m} \Pr\left(D^1(i,i+k)<r\mid D^{i-1}(1,1+k)<r\right) \label{eq:P_condP}
\end{align}
holds.

In the following, we show that the above conditional probability behaves oppositely to the Allan variance regardless of $k$, while the unconditional probability does not. This point is the key to understanding why $-C^m(r)$, the MSE and Allan variance show similar $\tau_A$ dependency.

Figure~\ref{fig:P_condP}(a) shows a logarithmic-scale color map of $\Pr(D^1(1,1+k)<r)$, which is the first term on the right-hand side of Eq. (\ref{eq:P_condP}), whereas Fig.~\ref{fig:P_condP}(b) shows that of $\Pr(D^1(2,2+k)<r \mid D^1(1,1+k)<r)$, corresponding to the conditional probability in the second term on the right-hand side of Eq.~(\ref{eq:P_condP}) for $i=2$.

Each probability was calculated for the LFF time series.
For $k=1$, corresponding to the lowest row in Fig.~\ref{fig:P_condP}(a) and the solid curve in Fig.~\ref{fig:P_condP}(c), the unconditioned probability $\Pr(D^1(1,2)<r)$ behaves opposite to the Allan variance (see Fig.~\ref{fig:LFF}(b)).
 However, for larger $k$, the behavior differs from that of the $k=1$ case, as shown in the dashed plot in Fig.~\ref{fig:P_condP}(c).

This can be understood as follows. 
First, the following equation holds:
\begin{align}
    &D^1(1,1+k)\notag\\
    =\,& 
    |\bar{s}^{(\tau_A)}(k+1)-\bar{s}^{(\tau_A)}(1)|\notag\\
    =\,&|\bar{s}^{(\tau_A)}(k+1)-\bar{s}^{(\tau_A)}(k) + \bar{s}^{(\tau_A)}(k) - \cdots + \bar{s}^{(\tau_A)}(2) - \bar{s}^{(\tau_A)}(1)|\notag\\
    =\,&\left|\sum_{j=2}^{k+1}\Delta\bar{s}^{(\tau_A)}(j)\right|. \label{eq:deltasum}
\end{align}
For $k=1$, whether $D^1(1,2) = |\Delta \bar{s}^{(\tau_A)}(2)|$ is smaller than $r$ is closely related to the Allan variance. 
If the Allan variance for a certain $\tau_A$ is smaller than that of another $\tau_A$, the $\Delta\bar{s}^{(\tau_A)}(l)$ distribution is biased toward the center, because the mean value of $\Delta\bar{s}^{(\tau_A)}(l)$ is zero by definition.
Consequently, the probability $D^1(1,2) = |\Delta \bar{s}^{(\tau_A)}(2)|<r$ is high for that $\tau_A$.
For larger $k$, the $\sum_{j=2}^{k+1}\Delta\bar{s}^{(\tau_A)}(j)$ distribution is affected by the time correlation of $\bar{s}^{(\tau_A)}(l)$ that the time series under study inherently has, and the Allan variance $\Var(\Delta\bar{s}^{(\tau_A)}(l))$ cannot predict the $\sum_{j=2}^{k+1}\Delta\bar{s}^{(\tau_A)}(j)$ distribution well, so $\Pr(D^1(1,1+k)<r)$ does not behave in a manner that is strongly correlated with the Allan variance.

Conversely, the conditional probability $\Pr(D^1(2,2+k)<r \mid D^1(1,1+k)<r)$ depicted in Fig.~\ref{fig:P_condP} (b) shows a similar trend to that of the Allan variance regardless of $k$, which we can also observe in Fig.~\ref{fig:P_condP}(d).
That is, $\Pr(D^1(2,2+k)<r \mid D^1(1,1+k)<r)$ is the main connection that explains the similarity between the MSE and Allan variance.
To examine this further, we discuss the variance of the corresponding quantity, rather than considering the probabilities in the following sections.

\subsubsection{Neighborhood-Likelihood-to-Variance-Relationship (NLVR) \label{sec:NLVR}}
In this section, we introduce an assumption called the Neighborhood-Likelihood-to-Variance-Relationship (NLVR) to connect the probability discussion to the variance discussion.
First, define $\Delta \Pr(\tau_A)$ as the difference between $\Pr(D^1(i,i+k)<r \mid D^{i-1}(1,1+k)<r)$ for $\tau_A$ and $\tau_A-1$. In the same way, $\Delta \Var(\tau_A)$ is the difference between $\Var(\bar{s}^{(\tau_A)}(i+k) - \bar{s}^{(\tau_A)}(i) \mid D^{i-1}(1,1+k)<r)$ for $\tau_A$ and $\tau_A-1$.
Please note that the conditional probability $\Pr(D^1(i,i+k)<r \mid D^{i-1}(1,1+k)<r)$ is the same as that in Eq.~(\ref{eq:P_condP}), and the absolute value of $\bar{s}^{(\tau_A)}(i+k) - \bar{s}^{(\tau_A)}(i)$, referred to in $\Delta \Var(\tau_A)$, is identical to $D^1(i,i+k)$ in the definition of $\Delta \Pr(\tau_A)$.

Then, to connect the probability with the variance, we assume the  Neighborhood-Likelihood-to-Variance-Relationship (NLVR), as follows:
\begin{align}
    \Delta\Pr(\tau_A)\Delta\Var(\tau_A) < 0. \label{iq:assumption}
\end{align}
The inequality (\ref{iq:assumption}) states that the signs of $\Delta\Pr(\tau_A)$ and $\Delta\Var(\tau_A)$ are always opposite.

We will explain the NLVR in more detail.
Both $\Pr(D^1(i,i+k)<r \mid D^{i-1}(1,1+k)<r)$ and $\Var(\bar{s}^{(\tau_A)}(i+k) - \bar{s}^{(\tau_A)}(i) \mid D^{i-1}(1,1+k)<r)$ are statistics of $\bar{s}^{(\tau_A)}(i+k) - \bar{s}^{(\tau_A)}(i)$ under the same condition of $D^{i-1}(1,1+k)<r$.
Considering the PDF of $\bar{s}^{(\tau_A)}(i+k) - \bar{s}^{(\tau_A)}(i)$, $\Pr(D^1(i,i+k)<r \mid D^{i-1}(1,1+k)<r)$ represents the area under the PDF in the range $[-r,~r]$.
In addition, when there is no condition, the mean of $\bar{s}^{(\tau_A)}(i+k) - \bar{s}^{(\tau_A)}(i)$ is zero by definition.
Here we also assume that the mean value of $\bar{s}^{(\tau_A)}(i+k) - \bar{s}^{(\tau_A)}(i)$ is close to zero under the condition $D^{i-1}(1,1+k)<r$.
When $\Delta\Pr(\tau_A)>0$, the distribution of $\bar{s}^{(\tau_A)}(i+k) - \bar{s}^{(\tau_A)}(i)$ is more centrally biased. 
Accordingly, when $\Delta\Pr(\tau_A)>0$, $\Delta\Var(\tau_A)$ is likely to be negative. 

\begin{figure}[htbp] 
    \centering
    \captionsetup{singlelinecheck=off,justification=ljustified}
    \includegraphics[width=\columnwidth]{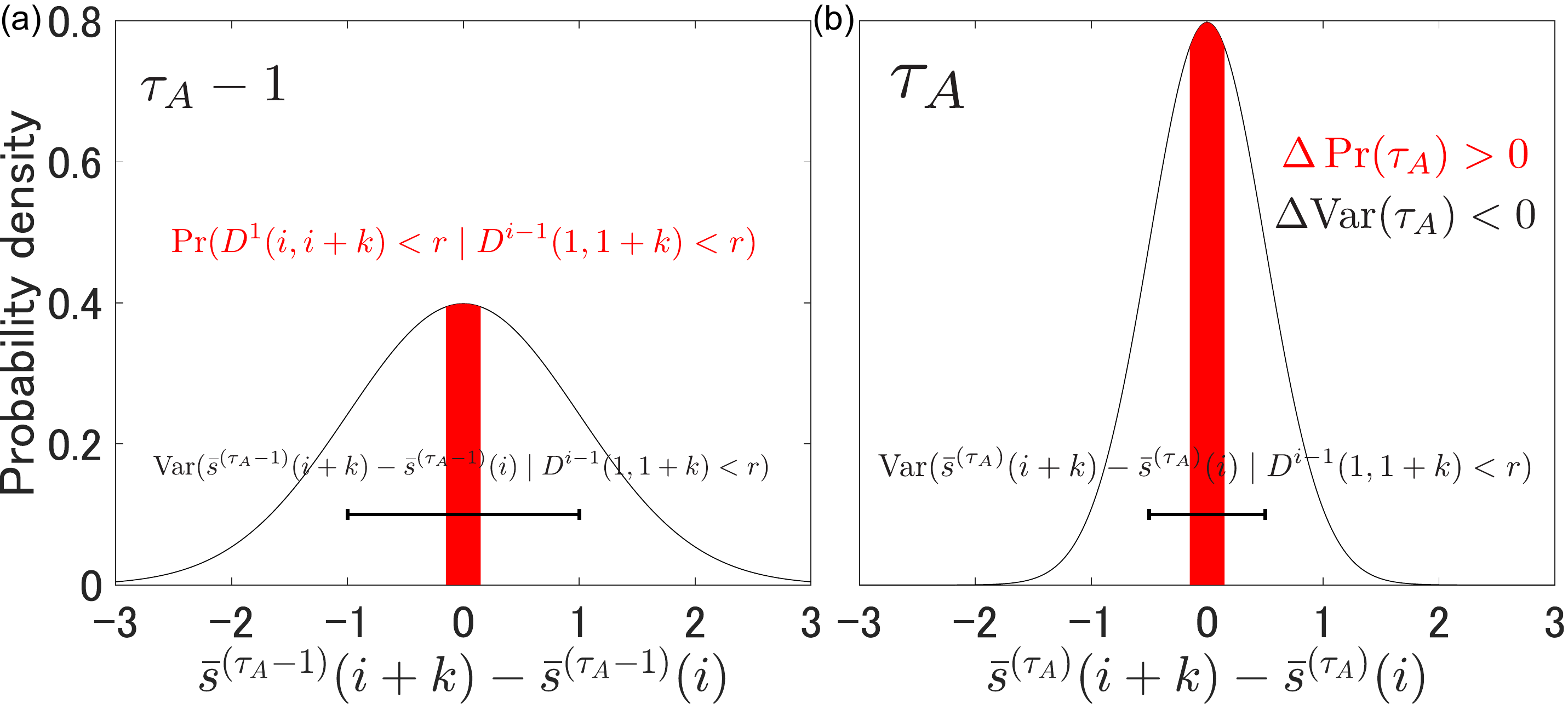}
    \caption{A visual representation of the NLVR. (a, b) $\bar{s}^{(\tau_A)}(i+k)-\bar{s}^{(\tau_A)}(i)$ distribution under the condition $D^{i-1}(1+k,1) < r$ for $\tau_A-1$ and $\tau_A$, respectively. The red-filled area and the the black bars represent $\Pr(D^1(i,i+k)<r \mid D^{i-1}(1,1+k)<r)$ and $\Var(\bar{s}^{(\tau_A)}(i+k) - \bar{s}^{(\tau_A)}(i) \mid D^{i-1}(1,1+k)<r)$, respectively. 
    }
    \label{fig:NLVR}
\end{figure}

Figure~\ref{fig:NLVR} schematically illustrates the concept of the NLVR.
The black curves in Figs. \ref{fig:NLVR} (a) and (b) represent the PDF of $\bar{s}^{(\tau_A)}(i+k)-\bar{s}^{(\tau_A)}(i)$ under the condition $D^{i-1}(1+k,1) < r$ for $\tau_A-1$ and $\tau_A$, respectively.
The red-filled areas and black bars show
$\Pr(D^1(i,i+k)<r \mid D^{i-1}(1,1+k)<r)$ and $\Var(\bar{s}^{(\tau_A)}(i+k) - \bar{s}^{(\tau_A)}(i) \mid D^{i-1}(1,1+k)<r)$, respectively.
$\Delta\Pr(\tau_A)$ and $\Delta\Var(\tau_A)$ are the difference in the red-filled areas and black bars between  Figs.~\ref{fig:NLVR}(b) and (a), respectively.
Please note that the plots shown here, based on Gaussian distributions, are for explanatory purposes only and are not obtained from a time series.
The validity of the NLVR is discussed in Sec.~\ref{sec:hypothesis}.

\subsubsection{Variance decomposition \label{sec:vardec}}
Here we further discuss the conditional variance $\Var(\bar{s}^{(\tau_A)}(i+k) - \bar{s}^{(\tau_A)}(i) \mid D^{i-1}(1,1+k)<r)$, instead of the conditional probability $\Pr(D^1(i,i+k)<r \mid D^{i-1}(1,1+k)<r)$ following the NLVR assumption, as mentioned in Sec.~\ref{sec:CIdec}.
By decomposing $\bar{s}^{(\tau_A)}(i+k) - \bar{s}^{(\tau_A)}(i)$ as
\begin{align}
    &\bar{s}^{(\tau_A)}(i+k) - \bar{s}^{(\tau_A)}(i) \notag\\
    =\,&\bar{s}^{(\tau_A)}(i+k)-\bar{s}^{(\tau_A)}(i+k-1)+\bar{s}^{(\tau_A)}(i+k-1)\notag\\ &- \bar{s}^{(\tau_A)}(i) + \bar{s}^{(\tau_A)}(i-1) - \bar{s}^{(\tau_A)}(i-1) \notag\\
    =\,&\Delta\bar{s}^{(\tau_A)}(i+k) - \Delta\bar{s}^{(\tau_A)}(i) + (\bar{s}^{(\tau_A)}(i+k-1)- \bar{s}^{(\tau_A)}(i-1)),
\end{align}
the conditional variance, 
$\Var(\bar{s}^{(\tau_A)}(i+k) - \bar{s}^{(\tau_A)}(i) \mid D^{i-1}(1,1+k)<r)$,
can be decomposed as follows:
\begin{align}
        &\Var(\bar{s}^{(\tau_A)}(i+k) - \bar{s}^{(\tau_A)}(i)\mid  D^{i-1}(1,1+k)<r) \notag \\
        =& \Var(\Delta\bar{s}^{(\tau_A)}(i+k)\mid  \cdots) \notag \\
         & +\Var(\Delta\bar{s}^{(\tau_A)}(i)\mid  \cdots) \notag \\
         & - 2 \Cov(\Delta\bar{s}^{(\tau_A)}(i+k), \Delta\bar{s}{(\tau_A)}(i)\mid  \cdots) \notag \\
         & + \Var(\bar{s}^{(\tau_A)}(i+k-1) -\bar{s}^{(\tau_A)}(i-1)\mid  \cdots) \notag \\
         & + 2 \Cov(\Delta\bar{s}^{(\tau_A)}(i+k), (\bar{s}^{(\tau_A)}(i+k-1) -\bar{s}^{(\tau_A)}(i-1))\mid \cdots) \notag \\
         & - 2 \Cov(\Delta\bar{s}^{(\tau_A)}(i), (\bar{s}^{(\tau_A)}(i+k-1) -\bar{s}^{(\tau_A)}(i-1))\mid \cdots) . \label{eq:VarDecompose}
\end{align}
Here, the condition $D^{i-1}(1,1+k)<r$ is abbreviated by the symbol $\cdots$ on the right-hand side of Eq.~(\ref{eq:VarDecompose}).

\vspace{2mm}
\noindent \textbf{[Overview of the variance decomposition]}

\begin{figure*}[htbp] 
\centering
\includegraphics[width=1.95\columnwidth]{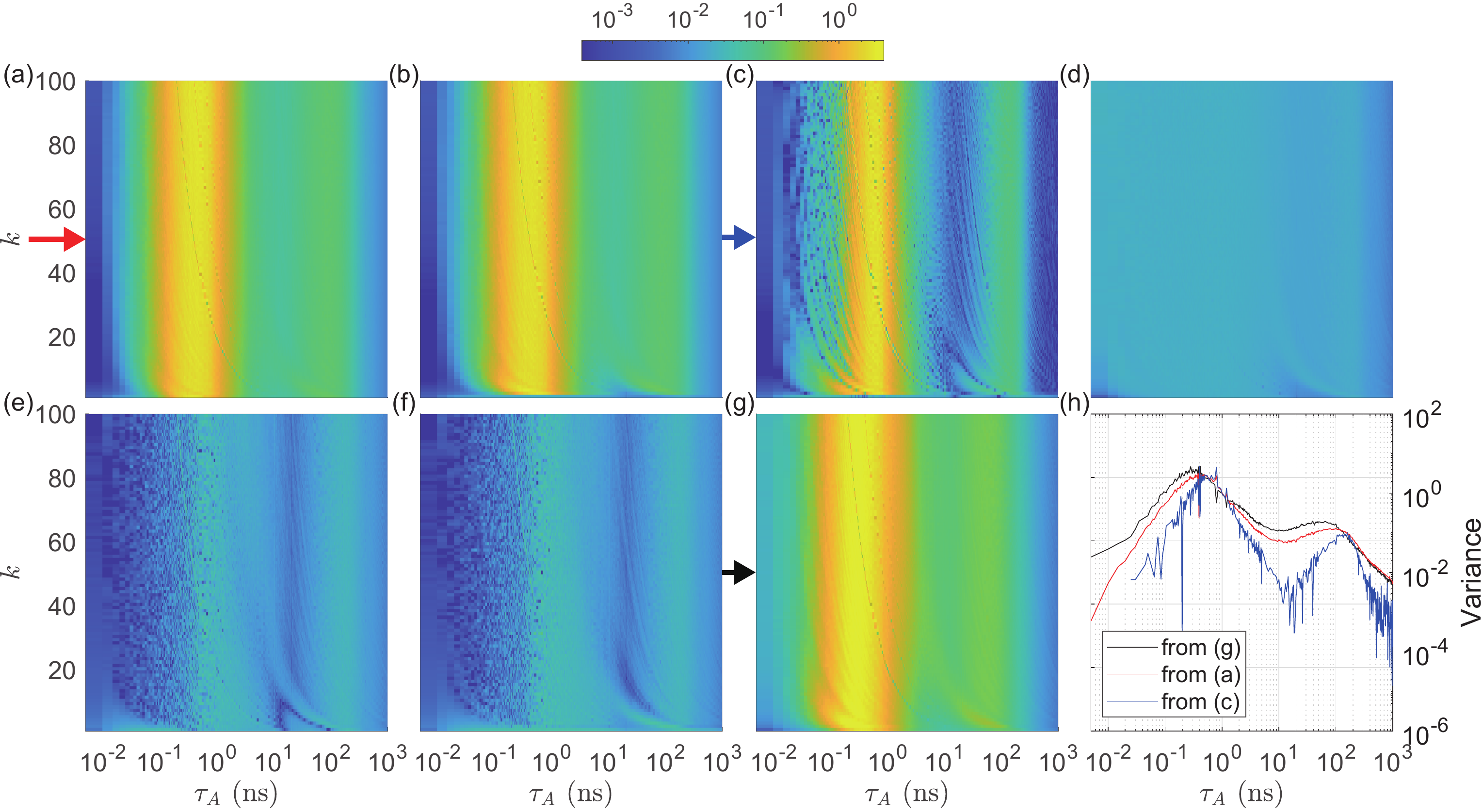}
\captionsetup{singlelinecheck=off,justification=ljustified}
\caption[
foo 
]{
Element-wise logarithmic-scale color maps of the conditional variance 
$\Var(\bar{s}^{(\tau_A)}(i+k) - \bar{s}^{(\tau_A)}(i)\mid  D^{i-1}(1,1+k)<r)$
given by Eq.~(\ref{eq:VarDecompose}). There are six terms of an LFF time series up to $k=100$: 
\begin{enumerate}[\textrm{(}a\textrm{)}]
    \item $\Var(\Delta\bar{s}^{(\tau_A)}(i+k)\mid  D^{i-1}(1,1+k)<r)$, 
    \item $\Var(\Delta\bar{s}^{(\tau_A)}(i)\mid  D^{i-1}(1,1+k)<r)$, 
    \item $2| \Cov(\Delta\bar{s}^{(\tau_A)}(i+k), \Delta\bar{s}^{(\tau_A)}(i)\mid  D^{i-1}(1,1+k)<r)|$, 
    \item $\Var(\bar{s}^{(\tau_A)}(i+k-1) -\bar{s}^{(\tau_A)}(i-1)\mid  D^{i-1}(1,1+k)<r)$, 
    \item $2 | \Cov(\Delta\bar{s}^{(\tau_A)}(i+k), (\bar{s}^{(\tau_A)}(i+k-1) -\bar{s}^{(\tau_A)}(i-1))\mid D^{i-1}(1,1+k)<r)|$ and 
    \item $2 | \Cov(\Delta\bar{s}^{(\tau_A)}(i), (\bar{s}^{(\tau_A)}(i+k-1) -\bar{s}^{(\tau_A)}(i-1))\mid D^{i-1}(1,1+k)<r)|$.
    \item The term before the above decomposition, which is $\Var(\bar{s}^{(\tau_A)}(i+k) - \bar{s}^{(\tau_A)}(i)\mid  D^{i-1}(1,1+k)<r)$.
    \item Cross-sectional profiles when $k=50$ from (a) (red curve), (c) (blue curve), and (g) (black curve) indicated by the arrows.
\end{enumerate}
Please note that for parts (c), (e), and (f) we plot the absolute value of the covariance, because covariance can be negative, and our color scale is logarithmic. However, $2\Cov(\Delta\bar{s}^{(\tau_A)}(i+k), \Delta\bar{s}^{(\tau_A)}(i)\mid  D^{i-1}(1,1+k)<r)$, shown in (c), is mostly positive. In (g), although the points with negative values are neglected, almost the entire curve is depicted.
}
\label{fig:Vars}
\end{figure*}

There are six terms on the right-hand side of Eq.~(\ref{eq:VarDecompose}). 
The logarithmic scale color map of each term is shown in Figs.~\ref{fig:Vars}(a)–(f) to determine which term has the greatest influence on the conditional variance. 
Figure~\ref{fig:Vars}(g) represents the summation of each term, which is the original conditional variance. 
Meanwhile, Fig.~\ref{fig:Vars}(h) show cross-sectional profiles when $k=50$ regarding the first term (Fig.~\ref{fig:Vars}(a)), the third term (Fig.~\ref{fig:Vars}(c)) and the total conditional variance (Fig.~\ref{fig:Vars}(g)).

As discussed shortly below, the first through third terms in Eq.~(\ref{eq:VarDecompose}) are dominant, as shown in Figs.~\ref{fig:Vars}(a)–(c), whereas they show a similar $\tau_A$ dependency with the Allan variance regardless of $k$, as indicated in Fig.~\ref{fig:Vars}(h).

Please note that we plotted the absolute values of the covariance terms, because the covariance can be negative. 
However, the $\tau_A$ dependency of $\Cov(\Delta\bar{s}^{(\tau_A)}(i+k), \Delta\bar{s}^{(\tau_A)}(i)\mid D^{i-1}(1,1+k)<r)$, shown in Figs.~\ref{fig:Vars}(c) and (h), is still similar to that of the Allan variance.
Conversely, the remaining panels (Figs.~\ref{fig:Vars}(d)–(f)) exhibit small values, regardless of $k$ and $\tau_A$. 

\begin{figure}[htbp] 
\centering
\includegraphics[width=\columnwidth]{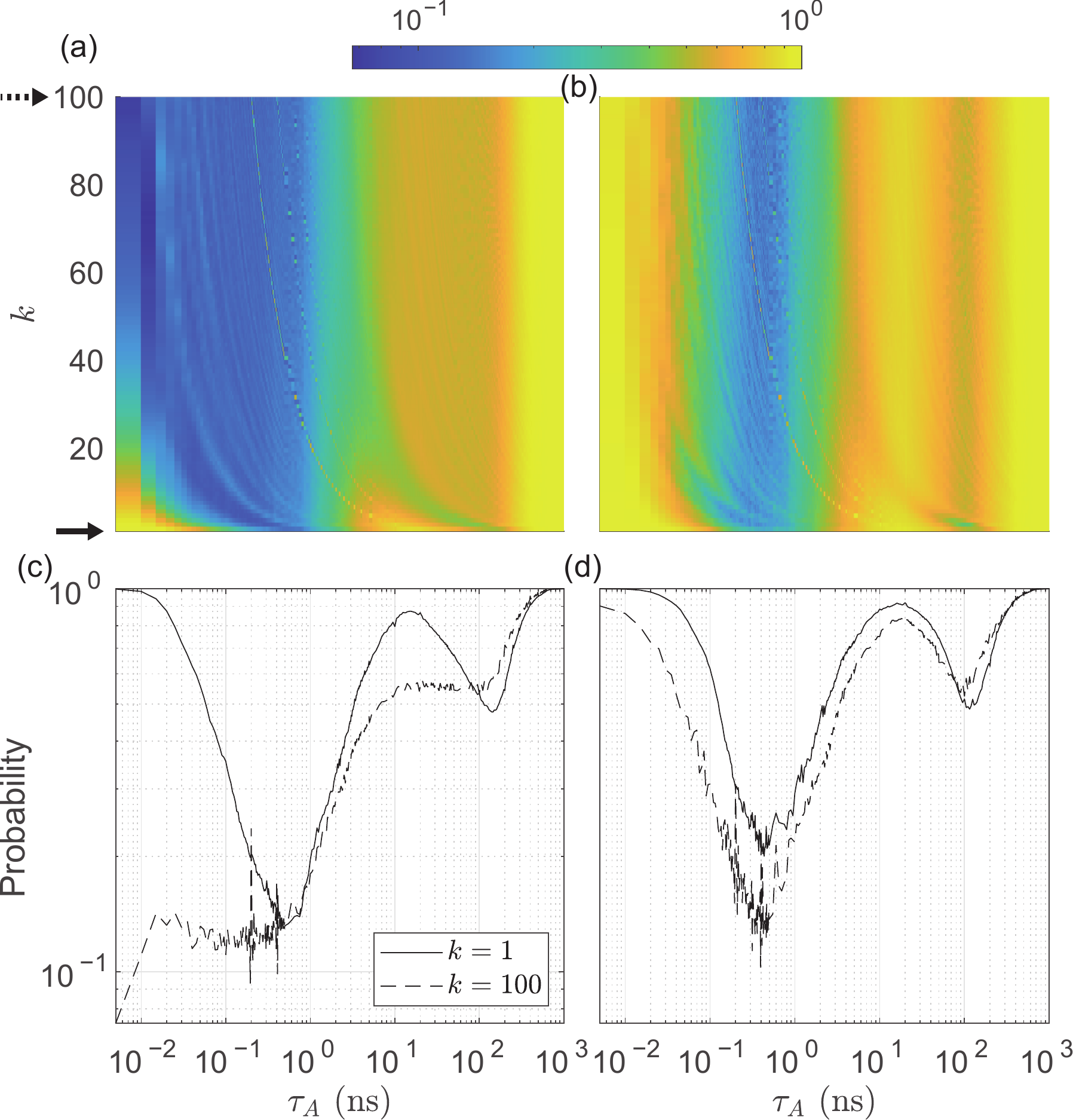}
\captionsetup{singlelinecheck=off,justification=ljustified}
\caption[
foo 
]{
(a,b)
Logarithmic scale color maps of 
\begin{enumerate}[\textrm{(}a\textrm{)}]
    \item $\Pr(D^1(1,1+k)<r)$ and 
    \item $\Pr(D^1(2,2+k)<r \mid D^1(1,1+k)<r)$
\end{enumerate}
for the LFF time series up to $k=100$, respectively.
(c,d)
One-dimensional plots of the two-dimensional data displayed in (a) and (b) along the two arrows ($k=1$ and $k=100$). 
Solid and dashed plots correspond to $k=1$ and $k=100$, respectively.}
\label{fig:P_condP}
\end{figure}

Therefore, we observe that the conditional variance 
$\Var(\bar{s}^{(\tau_A)}(i+k) - \bar{s}^{(\tau_A)}(i)\mid  D^{i-1}(1,1+k)<r)$, 
which is shown in Figs.~\ref{fig:Vars}(g) and (h), 
and the conditional probability 
$\Pr(|\bar{s}^{(\tau_A)}(i+k) - \bar{s}^{(\tau_A)}(i)|<r \mid D^{i-1}(1,1+k)<r)$, shown in Fig.~\ref{fig:P_condP}(b) for the case of $i=2$, 
exhibit the opposite $\tau_A$ dependency to the Allan variance from the NLVR.

\vspace{2mm}
\noindent \textbf{[The first and the second terms]}

The first and second terms are the conditional variances of $\Delta\bar{s}^{(\tau_A)}(i+k)$
and 
$\Delta\bar{s}^{(\tau_A)}(i)$, namely the conditional Allan variances, respectively. Although the condition somewhat affects the $\Delta\bar{s}^{(\tau_A)}(i+k)$
and 
$\Delta\bar{s}^{(\tau_A)}(i)$ distribution, the size relationship of the variance of $\Delta\bar{s}^{(\tau_A)}(i+k)$
and 
$\Delta\bar{s}^{(\tau_A)}(i)$ concerning $\tau_A$ is almost the same as the Allan variance.

\vspace{2mm}
\noindent \textbf{[The third term]}

\begin{figure}[htbp] 
\centering
\includegraphics[width=\columnwidth]{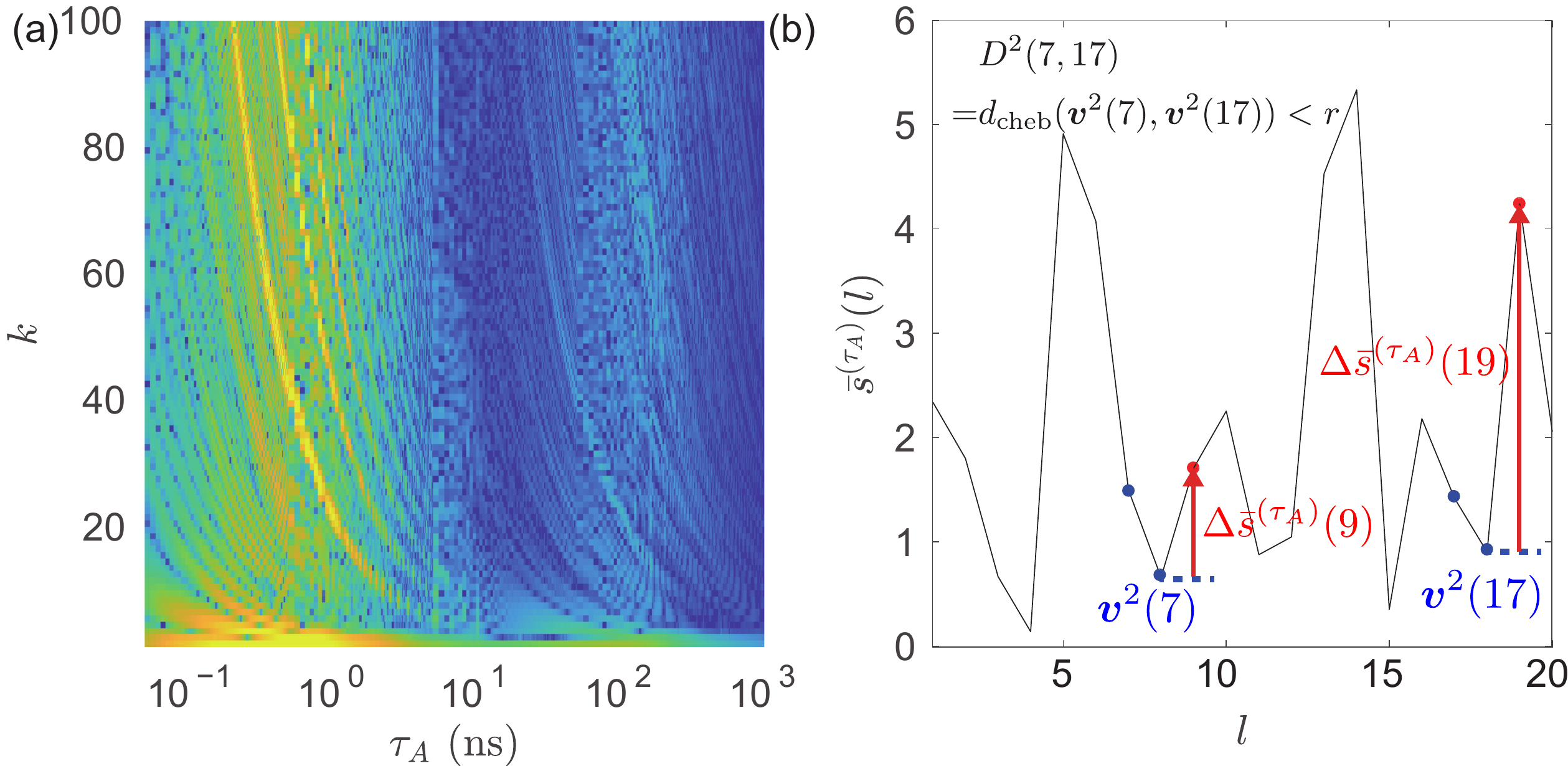}
\captionsetup{singlelinecheck=off,justification=ljustified}
\caption{
(a) The color map of the \textit{unconditional} version of the third term of the conditional variance (Eq.~(\ref{eq:VarDecompose}): $2|\Cov(\Delta\bar{s}^{(\tau_A)}(i+k), \Delta\bar{s}^{(\tau_A)}(i)|$. The color scale is the same as in Fig.~\ref{fig:Vars}.
(b) $\Delta\bar{s}^{(\tau_A)}(i+k)$ and $\Delta\bar{s}^{(\tau_A)}(i)$ under the condition of $D^{2}(7,7+10)<r$. Here, $i=3$ and $k=10$. The blue dots represent the two consecutive points which we can regard as $\bm{v}^2(7)$ and $\bm{v}^2(17)$ satisfying the condition $D^{2}(7,7+10)<r$.
The red dots denote $\bar{s}^{(\tau_A)}(9)$ and $\bar{s}^{(\tau_A)}(19)$ and the red arrows show $\Delta\bar{s}^{(\tau_A)}(9)$ and $\Delta\bar{s}^{(\tau_A)}(19)$.
}
\label{fig:covs}
\end{figure}

However, the third term is strongly influenced by the condition. 
In the absence of the condition, if $k$ is sufficiently large, 
there would be little correlation between 
$\Delta\bar{s}^{(\tau_A)}(i+k)$ and 
$\Delta\bar{s}^{(\tau_A)}(i)$; 
thus, the covariances are expected to be small. 
Fig.~\ref{fig:covs}(a) shows the unconditional covariance of $\Delta\bar{s}^{(\tau_A)}(i+k)$ and 
$\Delta\bar{s}^{(\tau_A)}(i)$, where the color scale is the same as in Fig.~\ref{fig:Vars}. Comparing to the conditional covariance plotted in Fig.~\ref{fig:Vars}(c), this covariance without the condition is smaller in most places.

By contrast, when the condition is satisfied, $i-1$ consecutive points up to 
$\bar{s}^{(\tau_A)}(i+k-1)$ and $\bar{s}^{(\tau_A)}(i-1)$ are close values. 
In that case, $\Delta\bar{s}^{(\tau_A)}(i+k)$ and $\Delta\bar{s}^{(\tau_A)}(i)$, i.e., the difference between these points and their consecutively following point, often have the same sign, and the distribution of their magnitudes follows the Allan variance. 

Figure~\ref{fig:covs}(b) shows an example of a pair of $\Delta\bar{s}^{(\tau_A)}(i+k)$ and 
$\Delta\bar{s}^{(\tau_A)}(i)$ satisfying the condition for $i=3$.
Because we can regard $\bm{v}^2(7)$ and $\bm{v}^2(17)$, denoted by the blue dots, as identical under the tolerance $r$, the red arrows, representing $\Delta\bar{s}^{(\tau_A)}(9)$ and 
$\Delta\bar{s}^{(\tau_A)}(19)$ are likely to be similar.
By the definition of covariance, if $\Delta\bar{s}^{(\tau_A)}(i+k)$ and 
$\Delta\bar{s}^{(\tau_A)}(i)$ are similar, $\Cov(\Delta\bar{s}^{(\tau_A)}(i+k),\Delta\bar{s}^{(\tau_A)}(i))$ is close to $\Var(\Delta\bar{s}^{(\tau_A)}(i+k))$, namely the Allan variance.
From these considerations, the conditional covariance behaves similarly to the Allan variance.

Please note that the form of the condition in the Fig.~\ref{fig:covs}(b), namely $D^2(7,17) < r$ is slightly different from $D^{i-1}(1,1+k)<r$. However, as we assume stationary, it is allowed to shift the indices. For example, in this case, we examined $\Delta\bar{s}^{(\tau_A)}(19)$ and $\Delta\bar{s}^{(\tau_A)}(9)$ under the condition of $D^2(7,17) < r$, and this corresponds to the shifting of indices by six. To be more specific, we investigated $\Delta\bar{s}^{(\tau_A)}(19) = \Delta\bar{s}^{(\tau_A)}(3+10+6)$ and $\Delta\bar{s}^{(\tau_A)}(3+6)$, under the condition of $D^2(7,17) < r = D^2(1+6,1+10+6)$, where $i = 3$ and $k=10$.

\vspace{2mm}
\noindent \textbf{[The fourth to the sixth terms]}

The fourth to sixth terms do not contribute significantly to the conditional variance, as mentioned earlier. 
This is due to the fact that $\bar{s}^{(\tau_A)}(i+k-1) -\bar{s}^{(\tau_A)}(i-1)$, which is involved in the variance or covariance in the fourth to sixth terms, is small when the condition $D^{i-1}(1,1+k)<r$ is satisfied. 
Once again, the condition implies a sort of similarity on the time scale of $i-1$ points, and when these points are similar, the variance of these terms containing the difference will be small.

\vspace{2mm}
\noindent \textbf{[Summary of the decomposition]}

Adding the above six terms together, we see that the conditional variance eventually behaves similarly to the Allan variance, as shown in Figs.~\ref{fig:Vars}(g) and (h).
Because the conditional variances and conditional probabilities are assumed by the NLVR to act in opposite directions for increasing scale factors, the conditional probabilities and $C^m(r)$, represented by the weighted average of the product of probability without conditions $\Pr(D^1(1,1+k)<r)$ and the conditional probabilities $\Pr(D^1(i,i+k)<r\mid D^{i-1}(1,1+k)<r)$, in turn, behave in the opposite direction for increasing scale factors for the Allan variance.

Although we observe the conditional variances and covariances from an LFF time series as an example, the discussion of behavior shown by the variances and covariances is valid for other time series to some extent.
In the discussion, we assumed strong stationarity of the signal $s(t)$ or its first-order difference to estimate probabilities and variances from the obtained time series. 
This is because $\Pr(D^m(i,j) <r)$ in Eq.~(\ref{eq:ij}) cannot be estimated from a single time series. 
Conversely, the NLVR assumption between probability and variance can also be applied to $\Pr(|\bar{s}^{(\tau_A)}(i)-\bar{s}^{(\tau_A)}(j)|<r)$ and $\Var(\bar{s}^{(\tau_A)}(i)-\bar{s}^{(\tau_A)}(j))$ and could be valid to a certain extent. 
Thus, similar behavior exhibited by the MSE and Allan variance should hold for a more general class of signals.


\subsection{\label{sec:MSEandAV} The MSE and Allan variance}
In Sec.~\ref{sec:CIandAV}, the connection between $C^m(r)$ and the Allan variance was discussed based on the connection between the conditional probability and variance. 
We can also apply this discussion to the connection between the MSE and Allan variance.

To simplify the discussion, we assume a sufficiently long (strongly stationary) time series, so that we can consider $C^m(r)$ to be the same as the expectation $\langle C^m(r)\rangle$. 
This is equivalent to assuming ergodicity. 

Under this condition,
\begin{align}
    &MSE \notag\\
    = &-\log \dfrac{C^{m+1}(r)}{C^m(r)} \notag\\
    = &-\log \dfrac{\sum_{k=1}^{M-m-1}(M-m-k)\Pr(D^{m+1}(1,1+k)<r)}{\sum_{k=1}^{M-m-1}(M-m-k)\Pr(D^m(1,1+k)<r)}. \label{eq:MSE1}
\end{align}
$\Pr(D^{m+1}(1,1+k)<r)$,
in the numerator of Eq.~(\ref{eq:MSE1}) can be decomposed similarly, as shown in Sec.~\ref{sec:CIandAV}, as follows:
\begin{align}
    &\Pr(D^{m+1}(1,1+k)<r) \notag \\
    = &\Pr(D^{1}(m+1,m+1+k)<r\mid D^m(1,1+k)<r)\notag\\
    &\Pr(D^{m}(1,1+k)<r).
\end{align}
Let $a_k = (M-m-k)\Pr(D^m(1,1+k)<r)$ and $b_k = \Pr(D^{1}(m+1,m+1+k)<r\mid D^m(1,1+k)<r)$.
Considering the equality $(M-m-k)\Pr(D^{m+1}(1,1+k)<r) = a_kb_k$, Eq.~(\ref{eq:MSE1}) can be rewritten as follows:
\begin{align}
  MSE &= -\log \dfrac{C^{m+1}(r)}{C^m(r)}\notag \\
      &= -\log \dfrac{\sum_{k=1}^{M-m-1}a_k b_k}{\sum_{k=1}^{M-m-1}a_k} \label{eq:MSE2}
\end{align}
Equation~(\ref{eq:MSE2}) states that the MSE is the negative logarithm of the weighted average of $b_k$. Because $b_k$ is the conditional probability, the behavior of the MSE can be explained by the conditional probability as well as $C^m(r)$.

More precisely, we must care about the expectation. $C^m(r)$ can be represented with probability when considering the expectation $\langle C^m(r)\rangle$.
The expectation of the MSE is
\begin{align}
    \langle MSE \rangle &= \left\langle -\log\dfrac{C^{m+1}(r)}{C^m(r)}\right\rangle \label{eq:bigexp}\\
    &\geq -\log \left\langle \dfrac{C^{m+1}(r)}{C^m(r)} \right\rangle \label{eq:jensen}.
\end{align}
From Eq.~(\ref{eq:bigexp}) to (\ref{eq:jensen}), we use the fact that the negative logarithm is convex, along with Jensen's inequality.
Eq.~(\ref{eq:jensen}) in general is not the same as
\begin{align}
    -\log \dfrac{\langle C^{m+1}(r)\rangle}{\langle C^m(r)\rangle} 
    = -\log \dfrac{\sum_{k=1}^{M-m-1}a_k b_k}{\sum_{k=1}^{M-m-1}a_k}.
\end{align}
However, it is noteworthy that Costa \textit{et al.}\cite{PhysRevE.71.021906} showed that the MSE values calculated theoretically, using probability, agree well for the WGN and $1/f$ noise cases.

\subsection{Connection between the probability and variance\label{sec:hypothesis}}
In Secs.~\ref{sec:CIandAV} and~\ref{sec:MSEandAV}, we discussed the connection between $C^m(r)$ or the MSE and Allan variance based on the assumption of NLVR. 
In this section, we discuss the extent to which NLVR is valid, and what happens if it is violated.

\vspace{2mm}
\noindent \textbf{[Considering independent identical distribution (i.i.d.)]}

For simplicity, we consider the case of independent identical distributions (i.i.d.). 
In this case, the conditional probability distribution of 
$(\bar{s}^{(\tau_A)}(i+k) - \bar{s}^{(\tau_A)}(i) \mid D^{i-1}(1,1+k)<r)$ 
is identical to the probability distribution of  
$(\bar{s}^{(\tau_A)}(i+k) - \bar{s}^{(\tau_A)}(i))$, 
which is independent of the condition $(D^{i-1}(1,1+k)<r)$.
The distribution does not depend on $k$ (see Appendix~\ref{sec:iid}). 
Therefore, it is sufficient to consider the distribution of 
$\bar{s}^{(\tau_A)}(i+1) - \bar{s}^{(\tau_A)}(i)$. 
In addition, the variances of $\bar{s}^{(\tau_A)}(i)$ and $\bar{s}^{(\tau_A)}(i+1) - \bar{s}^{(\tau_A)}(i)$ are $\Var(s(1))/\tau_A$ and $2\Var(s(1))/\tau_A$, respectively. 

Therefore, the Allan variance for an i.i.d. system is inversely proportional to $\tau_A$. 
For the WGN case, owing to the reproductive property, the PDF of $\bar{s}^{(\tau_A)}(i)$ and $\bar{s}^{(\tau_A)}(i+1) - \bar{s}^{(\tau_A)}(i)$ are also Gaussian, with variances $\Var(s(1))/\tau_A$ and $2\Var(s(1))/\tau_A$, respectively. In addition, because the mean value of $\bar{s}^{(\tau_A)}(i+1) - \bar{s}^{(\tau_A)}(i)$ is zero, the probability $|\bar{s}^{(\tau_A)}(i+1) - \bar{s}^{(\tau_A)}(i)|<r$ can be represented as follows:
\begin{align}
    &\Pr(|\bar{s}^{(\tau_A)}(i+1) - \bar{s}^{(\tau_A)}(i)|<r) \notag\\
    = &\sqrt{\dfrac{1}{2\pi2\sigma/\tau_A}}\int_{-r}^r \exp\left(-\dfrac{x^2}{2\cdot (2\sigma/\tau_A)^2}\right)dx,
\end{align}
where $\sigma^2 = \Var(s(1))$, and increases monotonically with increasing $\tau_A$.
Therefore, NLVR holds for all $\tau_A$.

The variance of $\bar{s}^{(\tau_A)}(i+1) - \bar{s}^{(\tau_A)}(i)$ continuously decreases when $\tau_A$ increases in the i.i.d. case. That is, $\Delta\Var(\tau_A)$ is negative for all $\tau_A$.
The cases in which NLVR does not hold are those in which the area under the PDF of $\bar{s}^{(\tau_A)}(i+1) - \bar{s}^{(\tau_A)}(i)$ in the range $[-r,~r]$ decreases with increasing $\tau_A$, which means $\Delta\Pr(\tau_A)$ is also negative.
This situation seems unlikely to occur when the i.i.d. process is unimodal. 
That is to say, the reduction of the variance means the shrinking of the distribution toward the center, which means an increase in the probabilities around the center. 

\vspace{2mm}
\noindent \textbf{[When NLVR is invalid]}

However, situations that violate NLVR are likely to occur when, for example, the original system exhibits a multimodal distribution.
It should be noted that $\bar{s}^{(\tau_A)}(i)$ of a multimodal distribution can have more peaks than the distribution of $s(t)$, owing to the effect of averaging. 
Thus, $\bar{s}^{(\tau_A)}(i+1) - \bar{s}^{(\tau_A)}(i)$ may also have many peaks. 

The number of peaks increases as $\tau_A$ increases; however, beyond a certain point, the number of peaks eventually decreases because the peaks fuse with each other. 
According to the central limit theorem, the distribution of $\bar{s}^{(\tau_A)}(i)$ and $\bar{s}^{(\tau_A)}(i+1) - \bar{s}^{(\tau_A)}(i)$ finally assumes shapes close to a Gaussian distribution in the i.i.d. case.
As the number of peaks increases, the area of the PDF of $\bar{s}^{(\tau_A)}(i+1) - \bar{s}^{(\tau_A)}(i)$ near the center is distributed to each peak, thus, $\Delta\Pr(\tau_A)$ is negative until the effects of this distribution and peaks merging become antagonistic.

\vspace{2mm}
\noindent \textbf{[Violation of NLVR by bimodal distributions]}

\begin{figure}[htbp] 
\centering
\includegraphics[width=\columnwidth]{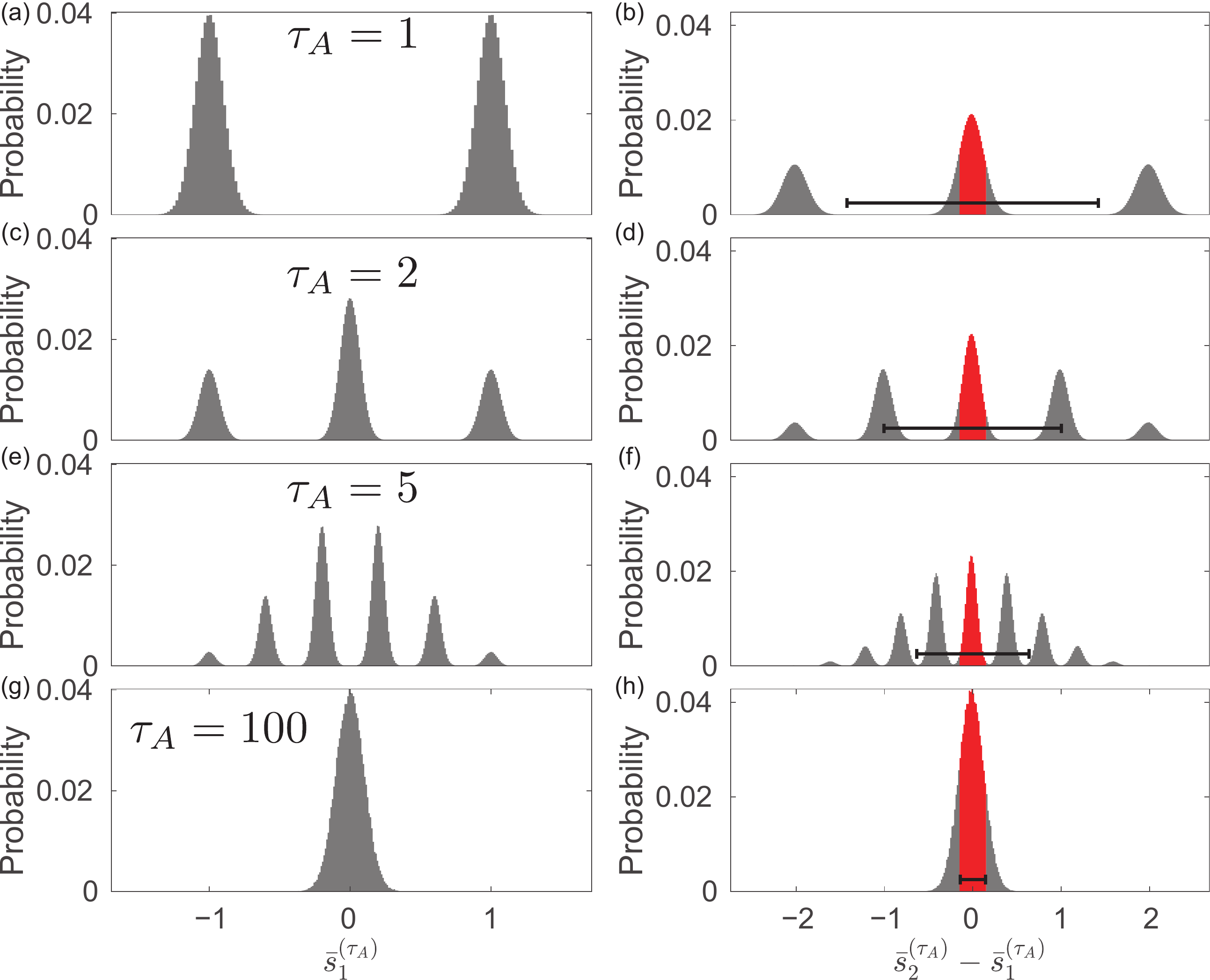}
\captionsetup{singlelinecheck=off,justification=ljustified}
\caption{
Example of histograms of a bimodal i.i.d. system. 
(a,c,e,g) Histograms of the coarse-grained time series for $\tau_A=1,~2,~5,~100$, respectively. 
(b,d,f,h) Histograms of the difference between adjacent coarse-grained points for $\tau_A=1,~2,~5,~100$, respectively. Red areas represent the range $[-r,~r]$. Black bars denote the standard deviation.}
\label{fig:bin_hist}
\end{figure}

Figure~\ref{fig:bin_hist} presents an example of this discussion. 
Figure~\ref{fig:bin_hist}(a) shows a histogram of the original bimodal i.i.d. system composed of two Gaussians centered around $\pm 1$.
Figures~\ref{fig:bin_hist}(c), (e), and (g) show the histograms of the coarse-grained time series for $\tau_A=1,~2,~5,~100$, respectively. 
Recall that coarse-graining means averaging over neighboring points.
Because the process is i.i.d., two consecutive points have a $50\,\%$ probability of coming from opposite sides of the origin. 
Thus, for $\tau_A = 2$ shown in Fig.~\ref{fig:bin_hist}(b), a third peak appears around the origin.
More precisely, the leftmost peak shown in Fig.~\ref{fig:bin_hist}(c) corresponds to the case in which two consecutive $s(t)$ points are negative, i.e., from the left peak in Fig.~\ref{fig:bin_hist}(a), which has a probability of $50\,\%\times 50\,\%=25\,\%$.
Similarly, the rightmost peak in Fig.~\ref{fig:bin_hist}(c) corresponds to two consecutive positive $s(t)$ points, and the center peak corresponds to either positive and negative, or negative and positive points.
The number of peaks increases up to a certain $\tau_A$, until they start to merge with each other. 
For a sufficiently large $\tau_A$, the distribution approaches a Gaussian distribution, as shown in Fig.~\ref{fig:bin_hist}(g).

Figures~\ref{fig:bin_hist}(b), (d), (f), and (h) show the histograms of the difference between adjacent coarse-grained points $\bar{s}^{(\tau_A)}(i+1)-\bar{s}^{(\tau_A)}(i)$ for $\tau_A=1,~2,~5,~100$, respectively. 
The red filled area represents the range $[-r,~r]$ and the black bar denotes the standard deviation of the distribution.
In fact, the distributions can be regarded as convolutions of the corresponding distribution of $\bar{s}^{(\tau_A)}(i)$. Similar to $\bar{s}^{(\tau_A)}(i)$, the number of peaks increases until a certain $\tau_A$ and then decreases. 
In such cases, $\Delta\Pr(\tau_A)$ is negative until a certain $\tau_A$, which can be observed from the reduction of the red area in Fig.~\ref{fig:bin_hist}(b)–(f), whereas $\Delta\Var(\tau_A)$ is always negative. Thus, the NLVR is violated.

\vspace{2mm}
\noindent \textbf{[Disagreement of MSE and Allan variance]}

\begin{figure}[htbp]  
    \centering
    \includegraphics[width=\columnwidth]{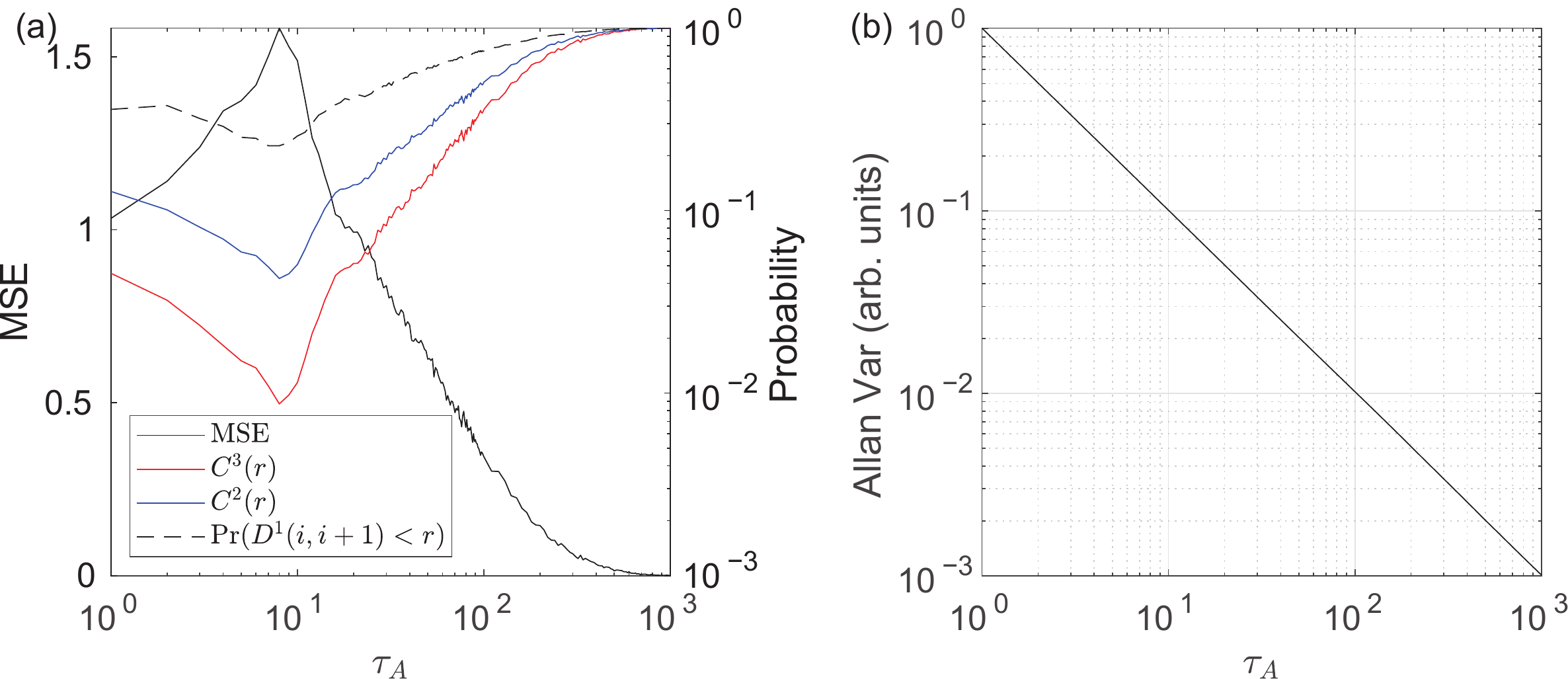} 
    \captionsetup{singlelinecheck=off,justification=ljustified}
    \caption{
Comparison of the MSE and Allan variance for the bimodal system shown in Fig.~\ref{fig:bin_hist}(a).
(a) The MSE, $C^m(r)$ and $\Pr(|\bar{s}^{(\tau_A)}(i+1)-\bar{s}^{(\tau_A)}(i)| < r)$ of a time series obtained from the system. 
The black, red, and blue curves represent the MSE, $C^3(r)$ and $C^2(r)$, respectively. 
The dashed black curve represents $\Pr(D^1(i,i+1)=|\bar{s}^{(\tau_A)}(i+1)-\bar{s}^{(\tau_A)}(i)| < r)$. 
(b) The Allan variance of the corresponding time series.
Herein, the MSE and the Allan variance clearly exhibit different $\tau_A$ dependencies.
}
    \label{fig:bin_mse}
\end{figure}

Plots of $\Pr(D^1(i,i+1)=|\bar{s}^{(\tau_A)}(i+1)-\bar{s}^{(\tau_A)}(i)| < r)$, the MSE, $C^m(r)$ and Allan variance of this bimodal i.i.d. system are presented in Fig.~\ref{fig:bin_mse}.
The dashed black line in Fig.~\ref{fig:bin_mse}(a) shows $\Pr(|\bar{s}^{(\tau_A)}(i+1)-\bar{s}^{(\tau_A)}(i)| < r)$. 
In contrast to the continuously decreasing Allan variance, $\Pr(|\bar{s}^{(\tau_A)}(i+1)-\bar{s}^{(\tau_A)}(i)| < r)$ decreases until $\tau_A = 8$ and then increases. $C^m(r)$ (red and blue curves) shows the same trend, and the MSE (black curve) inherits the inverted trend. 
Actually $\langle C^m(r) \rangle = \Pr(|\bar{s}^{(\tau_A)}(i+1)-\bar{s}^{(\tau_A)}(i)| < r)^m$ and $MSE\simeq -\log\Pr(|\bar{s}^{(\tau_A)}(i+1)-\bar{s}^{(\tau_A)}(i)| < r)$ (see Appendix~\ref{sec:iid}). 
In contrast, the Allan variance curve plotted in Fig.~\ref{fig:bin_mse}(b) shows $1/\tau_A$ dependency (note that the plot is logarithmically scaled), as discussed above.
Because the example under study here was an i.i.d. process, the Allan variance result may be considered the more reliable one, as it does not detect any specific time scales of relevance, whereas the MSE indicates a higher level of complexity for $\tau_A=8$ than at other scales.

From these discussions we can see that the NLVR is not always valid. 
For example, the condition may not hold when the original time series $s(t)$ contains a multimodal distribution. 
If the NLVR is violated, meaning that $\Delta\Pr(\tau_A)\Delta\Var(\tau_A)>0$ for a large number of $k$, $C^m(r)$, expressed as a weighted average of the probability, is expected to change in the same direction as the Allan variance. 
As a result, the Allan variance and MSE change with the opposite sign. 
In i.i.d. cases, $\Delta\Pr(\tau_A)$ and $\Delta\Var(\tau_A)$ are independent of $k$, so if NLVR does not hold for a specific $\tau_A$ and $k$, 
it is violated for all $k$. 

Therefore, in the range of $\tau_A$ where NLVR holds, the MSE and the Allan variance show similar $\tau_A$ dependence; 
in the range where the NLVR is violated, they behave oppositely to $\tau_A$. 
If the time series are time-correlated, as in dynamical systems, the NLVR may be violated by even more complex mechanisms.

\subsection{Empirical validation of the NLVR \label{sec:NLVRvalid}}
Finally, we examine the extent to which the NLVR is satisfied based on empirical data. 
Specifically, the change in conditional probability ($\Delta \Pr(\tau_A)$) and conditional variance ($\Delta \Var(\tau_A)$) are calculated for the time series studied in Sections~\ref{sec:simulation} and~\ref{sec:hypothesis} up to $k=100$. 

\begin{figure}[htbp]
\centering
\includegraphics[width=\columnwidth]{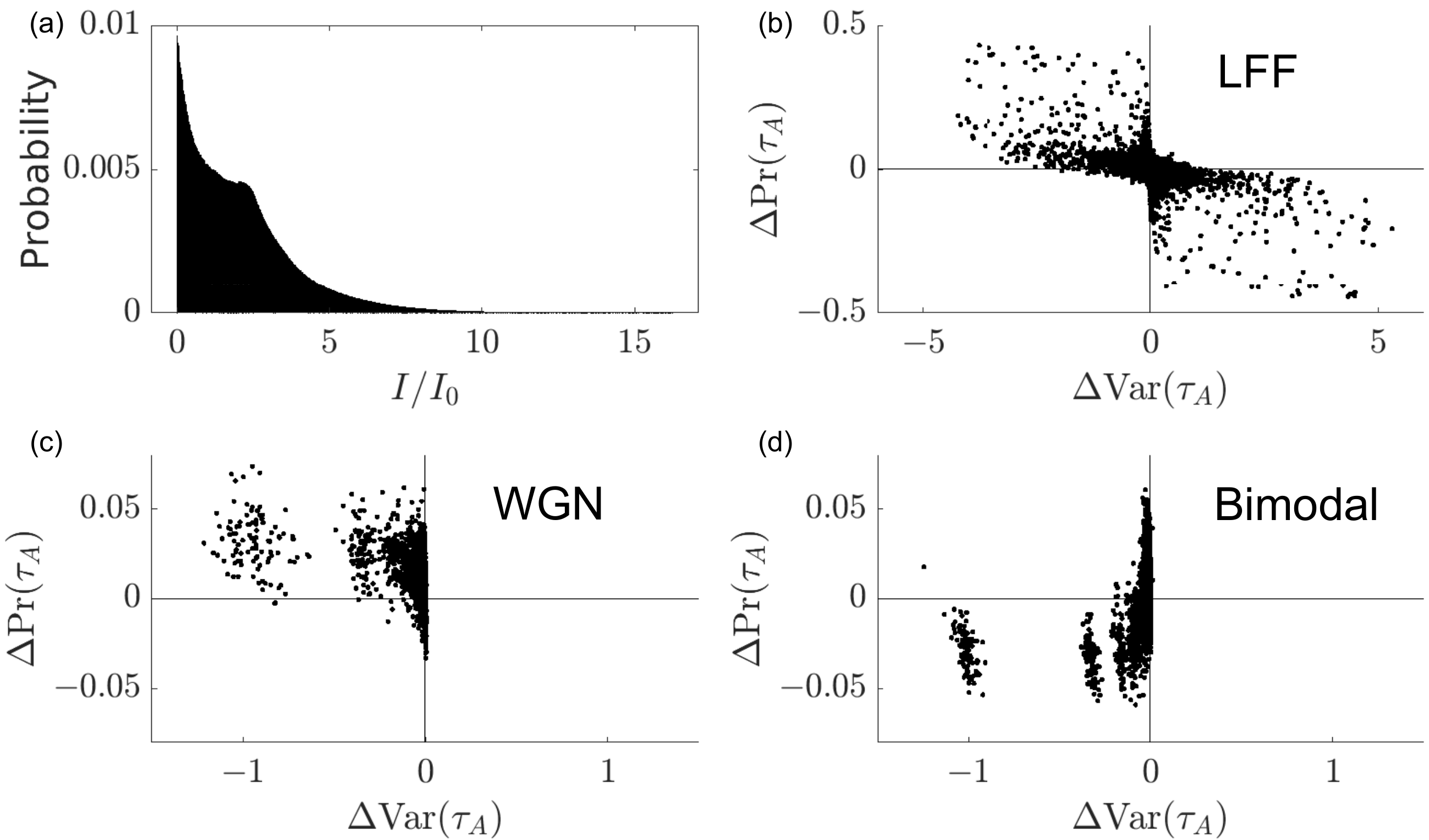}
\captionsetup{singlelinecheck=off,justification=ljustified}
\caption{
$\Delta\Pr(\tau_A)$ and $\Delta\Var(\tau_A)$ for $m=2$ and $i=2$, up to $k=100$.
(a) Histogram of the LFF time series. 
(b, c, d) Scatter plot of $\Delta\Pr(\tau_A)$ and $\Delta\Var(\tau_A)$ for (b) the LFF time series, (c) WGN time series, and (d) the bimodal time series. When the NLVR holds, the points in the scatter diagram should be in the second or fourth quadrants.
}
\label{fig:scatter}
\end{figure}

First, Fig.~\ref{fig:scatter}(a) shows a histogram of the intensity observed in the LFF time series. 
The intensity is normalized to that without optical feedback, i.e., the intensity in the single mode.\cite{Ohtsubo17} 
The histogram shown in Fig.~\ref{fig:scatter}(a) has a strong peak near the origin and a smaller peak near the normalized intensity of 3. 
That is, the probability distribution exhibits a somewhat multimodal distribution, which may cause a violation of the NLVR.

Figures \ref{fig:scatter}(b), (c) and (d) show the scatter plot of ($\Delta \Pr(\tau_A), \Delta \Var(\tau_A)$) for LFF, WGN, and bimodal distribution, respectively, as explained below. 
If the NLVR holds, a positive $\Delta \Pr(\tau_A)$ means a 
negative $\Delta \Var(\tau_A)$ and 
a negative $\Delta \Pr(\tau_A)$ means a positive $\Delta \Var(\tau_A)$.  
That is to say, the points in the scatter diagram should be in the second or fourth quadrants.

From Fig.~\ref{fig:scatter}(b), almost all the sampling points are concentrated in the second and fourth quadrants, i.e., in regions where the signs of $\Delta\Pr(\tau_A)$ and $\Delta\Var(\tau_A)$ are opposite. 
That is to say, even though the probability distribution contains multimodality, this LFF system mostly does not violate the NLVR. 
From these observations, we speculate that the peaks of the distribution would need to be more separated to lead to a disagreement of the Allan variance and MSE. 

Similarly, Fig.~\ref{fig:scatter}(c) shows $\Delta\Pr(\tau_A)$ and $\Delta\Var(\tau_A)$ for the WGN time series. 
As discussed above, $\Delta\Pr(\tau_A)$ should always be positive, whereas $\Delta\Var(\tau_A)$ is always negative, i.e., every point should be in the second quadrant. 
However, the distribution calculated from the time series is not precisely the Gaussian distribution, so some points are in the third quadrant.

Finally, Fig.~\ref{fig:scatter}(d) shows the same scatter plot for the bimodal case. 
In contrast to the LFF and WGN cases, many points are in the third quadrant in violation of the NLVR. 
This was expected, as we constructed this case specifically as a counter-example.

\subsection{Computation of the MSE and Allan variance \label{sec:comp}}
As introduced in Sec.~\ref{sec:MSE}, the definition of MSE involves probabilities, whereas that of the Allan variance is based on the variability of the data under study. 
Section~\ref{sec:hypothesis} reveals the common underlying mechanism from an information-theoretic viewpoint despite the seemingly different definitions of the statistical measures of multiscale dynamics. 
In this section, we discuss the difference from the viewpoint of computation between the MSE and Allan variance.

For the MSE, calculating $C^m(r)$ is a computationally demanding task. 
Computing Eq.~(\ref{eq:Cmi}) for all $i$ requires $O(N^2)$ calculations, as all pairs of embedded vectors are compared. 
Some implementations also require $O(N^2)$ memory to store all the calculated Chebyshev distances. 
Thus, the MSE evaluates the details of the probability distribution at a high computational cost. 

In contrast, the Allan variance calculation requires only $O(N)$ computations and memory.
The Allan variance does not consider the probability distribution; it depends only on the differences $\Delta\bar{s}^{(\tau_A)}(l)$ of successive coarse-grained points $\bar{s}^{(\tau_A)}(l)$.
As discussed in Sec.~\ref{sec:hypothesis}, the MSE for the i.i.d. process depends on the distribution, whereas the Allan variance does not. 

Despite being computationally cheaper, it is not obvious that the Allan variance has any disadvantages when compared to the MSE for extracting multiscale features. 
The Allan variance is a statistical measure that is similar but not identical to the MSE, and that quantifies slightly different aspects of the time series. 

In the literature, the range of applications of the MSE is versatile, such as bearing fault detection\cite{WANG2020107574} and sleep level qualification.\cite{6165354} 
However, the MSE suffers from severe computational difficulties as discussed above. 
Concerning the similar properties of the MSE and Allan variance, as well as the computationally lightweight nature of Allan variance, the extension of the Allan variance to real-time applications, such as bearing fault detection or prediction of epilepsy from electroencephalography (EEG), would be an interesting future topic.

\section{\label{sec:conclusion} Conclusion}
In this study, we examined the similarities shown by the multiscale statistics the MSE and Allan variance, and discussed the underlying mechanisms through an information-theoretic analysis. 
It is noteworthy that although the apparent definitions of the MSE and Allan variance are significantly different, they show a similar behavior for a wide range of time-series data.
We experimentally confirmed the similar properties of the MSE and Allan variance observed in LFF in chaotic lasers and physiological heartbeat data, as well as white Gaussian and $1/f$ noise.
The connection can be understood by decomposing the conditional probabilities in the MSE and extracting the dominant contributions. 
We derived a condition which must be satisfied for the MSE and Allan variance to exhibit similar tendencies via a discussion of conditional probabilities. 
Then, we artificially constructed a random sequence that violates the condition, leading to inconsistent MSE and Allan variance behavior. 
We also quantitatively demonstrated that the aforementioned LFF and heartbeat, which are physically plausible systems, mostly satisfy the condition.

Finally, we discussed future research topics.
Using Allan variance instead of the MSE may lead to more computationally lightweight applications that are suitable for real-time tasks.
In addition, there is a possibility of integrating further developments that have been devised for the MSE \cite{e17053110} and Allan variance.\cite{FSTABILITY}

Furthermore, more research on the theoretical foundations of coarse-graining and the MSE is needed. 
The MSE research to date has focused mainly on its application as a statistical tool, and there has been little research on its theoretical foundations. 
The relationship between the dynamics of a coarse-grained time series and those of the original time series is unclear. 
Coarse-graining can be regarded as a combination of a moving-average filter and downsampling; however, according to a previous study,\cite{embedology} a linear filter applied to the original time series during Takens' embedding preserves its topological properties. 
From this theorem, it may be possible to discuss the theoretical basis of coarse-graining from the viewpoint of dynamical invariants, including entropies.
Meanwhile, the theoretical foundations of the MSE should be more complicated, as the MSE shares the tolerance $r$ for all scales. As pointed out by Humeau-Heurtier,\cite{e17053110} more and more embedded vectors may be regarded as neighbors of each other as the scale factor $\tau_A$ increases, owing to the reduction of the variance of the coarse-grained time series.
Notably, Costa \textit{et al.} \cite{PhysRevE.71.021906}, the original proposers of the MSE, pointed out that the variance changes induced by coarse-graining are related to the temporal structures of the original time series.
We may need a framework that allows us to connect the dynamical invariants at each time scale.

\section*{Acknowledgments }
This work was supported in part by the CREST Project (JPMJCR17N2) funded by the Japan Science and Technology Agency and Grants-in-Aid for Scientific Research (JP20H00233), and  Transformative Research Areas (A) (JP22H05197) funded by the Japan Society for the Promotion of Science (JSPS). AR is supported by JSPS as an International Research Fellow.

\appendix
\section{Independent identical distributions case \label{sec:iid}}
In this section, we discuss the properties of conditional probability 
$\Pr(D^1(i,i+k)<r \mid D^{i-1}(1,1+k)<r)$ and conditional variance 
$\Var(\bar{s}^{(\tau_A)}(i+k) - \bar{s}^{(\tau_A)}(i) \mid D^{i-1}(1,1+k)<r)$ 
for i.i.d. systems.

Let $s(t)$ be the i.i.d. time series under study. We can regard $s(t)$ as a sequence of random variables.
Obviously, the coarse-grained time series defined as
\begin{align}
    \bar{s}^{(\tau_A)}(l) &= \dfrac{1}{\tau_A}\sum_{t=(l-1)\tau_A+1}^{l\tau_A} s(t)
\end{align}
is also an i.i.d. sequence of random variables. 
To simplify the symbols, let the random variable $Y_l$ be $\bar{s}^{(\tau_A)}(l)$. 
We now define the PDF of $Y_l = \bar{s}^{(\tau_A)}(l)$ as $g^{(\tau_A)}(y_l)$.
Please note that the function itself is independent of $l$.

The distribution of $\bar{s}^{(\tau_A)}(i+k) - \bar{s}^{(\tau_A)}(i)$ under the condition $D^{i-1}(1,1+k)<r$ is then computed.
For visibility, we define the random variable $Z_k$ as follows:
\begin{align}
    Z_k &= \bar{s}^{(\tau_A)}(i+k) - \bar{s}^{(\tau_A)}(i).
\end{align}
Let $h_k^{(\tau_A)}(z_k)$ be the PDF of $Z_k$ under the condition of $D^{i-1}(1,1+k)<r$.
$h_k^{(\tau_A)}(z_k)$ can be represented by $g^{(\tau_A)}(y_l)$ as follows:
\begin{align}
    h_k^{(\tau_A)}(z_k) &= \dfrac{d}{dz_k}\Pr(Z_k<z_k \mid D^{i-1}(1,1+k)<r)\notag\\
    &= \dfrac{d}{dz_k} \dfrac{\Pr(Z_k<z_k \land D^{i-1}(1,1+k)<r)}{\Pr(D^{i-1}(1,1+k)<r)} \notag\\
    &=\dfrac{d}{d z_k} \dfrac{\int_{T_{k,1}^{(\tau_A)}\land T_{k,2}^{(\tau_A)}}\prod_{j=1}^{i+k}g^{(\tau_A)}(y_j)dy_j}{\int_{T_{k,2}^{(\tau_A)}}\prod_{j=1}^{i+k}g^{(\tau_A)}(y_j)dy_j}, \label{eq:Ccond_int}
\end{align}
where
\begin{align}
    T_{k,1}^{(\tau_A)} &= \{(y_1,y_2,\cdots,y_{i+k})\mid y_{i+k}-y_i < z_k\}, \label{eq:Ccond1}\\
    T_{k,2}^{(\tau_A)} &= \{(y_1,y_2,\cdots,y_{i+k})\mid D^{i-1}(1,1+k) < r\}. \label{eq:Ccond2}
\end{align}
Here we used the fact that the joint PDF of i.i.d. random variables is equal to the product of PDFs of each random variable.
Because the conditions in Eqs.~(\ref{eq:Ccond1}) and (\ref{eq:Ccond2}) refer to different variables, the integral in the numerator of Eq.~(\ref{eq:Ccond_int}) can be separated into the integrals of variables $y_i$ and $y_{i+k}$, and the other terms, as follows:
\begin{align}
    &\int_{T_{k,1}^{(\tau_A)}\land T_{k,2}^{(\tau_A)}}\prod_{j=1}^{i+k}g^{(\tau_A)}(y_j)dy_j \notag\\
    = &\int_{\tilde{T}_{k,1}^{(\tau_A)}}g^{(\tau_A)}(y_i)g^{(\tau_A)}(y_{i+k})dy_idy_{i+k}\int_{\tilde{T}_{k,2}^{(\tau_A)}}\prod_{\substack{j=1\\j\neq i}}^{i+k-1}g^{(\tau_A)}(y_j)dy_j,
\end{align}
where
\begin{align}
    &\tilde{T}_{k,1}^{(\tau_A)}= \{(y_i,y_{i+k})\mid y_{i+k}-y_i < z_k\}, \label{eq:tildeCcond1}\\
    &\tilde{T}_{k,2}^{(\tau_A)}=\{(y_1,\cdots,y_{i-1},y_{i+1},\cdots,y_{i+k-1})\mid D^{i-1}(1,1+k) < r\}. \label{eq:tildeCcond2}
\end{align}
Here, different from Eqs.~(\ref{eq:Ccond1}) and (\ref{eq:Ccond2}), there is no variable overlap between Eqs.~(\ref{eq:tildeCcond1}) and (\ref{eq:tildeCcond2}).
Similarly, the denominator can also be decomposed as follows:
\begin{align}
    &\int_{T_{k,2}^{(\tau_A)}}\prod_{j=1}^{i+k}g^{(\tau_A)}(y_j)dy_j \notag\\
    = &\int_{\mathbb{R}^2}g^{(\tau_A)}(y_i)g^{(\tau_A)}(y_{i+k})dy_idy_{i+k}\int_{\tilde{T}_{k,2}^{(\tau_A)}}\prod_{\substack{j=1\\j\neq i}}^{i+k-1}g^{(\tau_A)}(y_j)dy_j \notag\\
    = &\int_{\tilde{T}_{k,2}^{(\tau_A)}}\prod_{\substack{j=1\\j\neq i}}^{i+k-1}g^{(\tau_A)}(y_j)dy_j.
\end{align}
Here we used the fact that
\begin{align}
    \int_{\mathbb{R}^2}g^{(\tau_A)}(y_i)g^{(\tau_A)}(y_{i+k})dy_idy_{i+k} = 1.
\end{align}
As a result, Eq.~(\ref{eq:Ccond_int}) can be reduced to
\begin{align}
    h_k^{(\tau_A)}(z_k) &= \dfrac{d}{d z_k} \int_{\tilde{T}_{k,1}^{(\tau_A)}} g^{(\tau_A)}(y_i)g^{(\tau_A)}(y_{i+k})dy_idy_{i+k}. \label{eq:Ccond_reduced}
\end{align}
Equation~(\ref{eq:Ccond_reduced}) is the PDF of $Z_k$ without any conditions. 
In conclusion, the condition $D^{i-1}(1,1+k) < r$ does not matter in i.i.d. cases. 
In addition, the above calculation does not depend on $k$ and $i$ except for $k=1$, as $i+k-1 = i$.
Here $y_i$ appears in both $T^{(\tau_A)}_{k,1}$ and $T^{(\tau_A)}_{k,2}$, so we cannot divide the integral in the same manner.
However, the same conclusion can be derived for $k=1$.
Here we introduce a variable transformation, as follows:
\begin{align}
    u_j = 
    \begin{cases}
        y_{j+1} -y_j & (j=1,2,\cdots,i) \\
        y_{i+1} & (j=i+1).
    \end{cases}
\end{align}
Using $u_j$, the integral of the numerator in Eq.~(\ref{eq:Ccond_int}) can be written as follows:
\begin{align}
    \int_{-\infty}^{\infty}du_{i+1}\int_{-\infty}^{z_k}du_i\int_{-r}^r \prod_{j=1}^{i+1}g^{(\tau_A)}\left(2u_{i+1}-\left(\sum_{l=j}^{i+1}u_l\right)\right)\prod_{j=1}^{i-1}du_j. \label{eq:k_1_nu}
\end{align}
Similarly, the integral in the denominator in Eq.~(\ref{eq:Ccond_int}) is
\begin{align}
        \int_{-\infty}^{\infty}du_{i+1}\int_{-\infty}^{\infty}du_i\int_{-r}^r \prod_{j=1}^{i+1}g^{(\tau_A)}\left(2u_{i+1}-\left(\sum_{l=j}^{i+1}u_l\right)\right)\prod_{j=1}^{i-1}du_j. \label{eq:k_1_de}
\end{align}
Because the only difference between Eqs.~(\ref{eq:k_1_nu}) and (\ref{eq:k_1_de}) is the range of the integral for $u_i$, we can cancel the integrals for $u_1,u_2,\cdots,u_{i-1}$, and the remaining integrals for the numerator and denominator are
\begin{align}
    \int_{-\infty}^\infty du_{i+1} \int_{-\infty}^{z_k}du_i ~g^{(\tau_A)}(u_{i+1})g^{(\tau_A)}(u_{i+1}-u_i)
\end{align}
and
\begin{align}
    \int_{-\infty}^\infty du_{i+1} \int_{-\infty}^{\infty}du_i ~g^{(\tau_A)}(u_{i+1})g^{(\tau_A)}(u_{i+1}-u_i) = 1,
\end{align}
respectively.
Consequently, the resulting PDF is
\begin{align}
    h_1^{(\tau_A)}(z_1) &= \dfrac{d}{dz_1}\int_{-\infty}^\infty du_{i+1} \int_{-\infty}^{z_1}du_i ~g^{(\tau_A)}(u_{i+1})g^{(\tau_A)}(u_{i+1}-u_i) \notag\\
    &=\dfrac{d}{d z_1} \int_{\tilde{T}_{1,1}^{(\tau_A)}} g^{(\tau_A)}(y_i)g^{(\tau_A)}(y_{i+1})dy_idy_{i+1}.
\end{align}
This is equivalent to the PDF of $Z_1$ without any conditions.
Summarizing the results thus far, $h_k^{(\tau_A)}$ is the same as the PDF of $Z_k$ without any conditions for all $k$.
In addition, the calculation does not depend on $i$.
Consequently, it is sufficient to discuss the distribution of $\bar{s}^{(\tau_A)}(2)-\bar{s}^{(\tau_A)}(1)$ in i.i.d. cases.

From the above results, $\Pr(D^m(1,1+k)<r)$ can be expressed as follows:
\begin{align}
   &\Pr(D^m(1,1+k)<r) \notag \\
    = &\Pr(D^1(1,1+k)<r)\notag\\&\prod_{i=2}^m \Pr\left(D^1(i,i+k)<r\mid D^{i-1}(1,1+k)<r\right) \label{eq:iid_dec1}\\
    = &\Pr(D^1(1,1+k)<r)\prod_{i=2}^m \Pr\left(D^1(i,i+k)<r\right) \label{eq:iid_dec2}\\
    = &\Pr(D^1(1,1+k)<r)^m \label{eq:iid_dec3}\\
    = &\Pr(D^1(1,2)<r)^m. \label{eq:iid_dec4}
\end{align}
Here, we obtain Eq.~(\ref{eq:iid_dec1}) in the same manner as Eq.~(\ref{eq:P_condP}).
From Eq.~(\ref{eq:iid_dec1}) to (\ref{eq:iid_dec2}), we ignored the condition term, as discussed above. From Eq.~(\ref{eq:iid_dec2}) to(\ref{eq:iid_dec3}) and (\ref{eq:iid_dec4}), we used the fact that the distribution of $\bar{s}^{(\tau_A)}(i+k)-\bar{s}^{(\tau_A)}(i)$ is independent of $i$ and $k$.
Consequently, $\langle C^m(r) \rangle = \Pr(D^1(1,2)<r)^m$, and $MSE\simeq -\log\Pr(D^1(1,2)<r)$ holds.

It is noteworthy that the variances of $\bar{s}^{(\tau_A)}(1)$ and $\bar{s}^{(\tau_A)}(2)-\bar{s}^{(\tau_A)}(1)$ are $1/\tau_A$ and $2/\tau_A$ respectively, with regard to the variance of $s(1)$, as
\begin{align}
    \Var(\bar{s}^{(\tau_A)}(1))
    &=\Var\left(\dfrac{1}{\tau_A}\sum_{t=1}^{\tau_A}s(t)\right)\notag \\
    &= \dfrac{1}{\tau_A^2}\sum_{t=1}^{\tau_A}\Var(s(t))\notag \\
    &= \dfrac{1}{\tau_A}\Var(s(1)),
\end{align}
and
\begin{align}
    &\Var(\bar{s}^{(\tau_A)}(2)-\bar{s}^{(\tau_A)}(1)) \notag\\
    =\,&\Var(\bar{s}^{(\tau_A)}(2)) 
    + \Var(\bar{s}^{(\tau_A)}(1)) \notag\\
    =\,& \dfrac{2}{\tau_A} \Var(s(1))
\end{align}
hold because $s(t)$ and $\bar{s}^{(\tau_A)}(l)$ are i.i.d.

\bibliography{aipsamp}

\end{document}